\documentclass{aa}
\usepackage{graphicx}
\usepackage{natbib}
\usepackage{txfonts}
\begin{document}
\title{Rotation- and temperature-dependence of stellar latitudinal
  differential rotation\thanks{Based on observations carried out at
    the European Southern Observatory, Paranal and La Silla, PIDs
    68.D-0181, 69.D-0015, 71.D-0127, 72.D-0159, 73.D-0139, and on
    observations collected at the Centro Astron\'omico Hispano
    Alem\'an (CAHA) at Calar Alto, operated jointly by the Max-Planck
    Institut f\"ur Astronomie and the Instituto de Astrof\'isica de
    Andaluc\'ia (CSIC)}}


\author{A. Reiners\inst{1,2}\thanks{Marie Curie International Outgoing Fellow}
}


\institute{ Astronomy Department, 521 Campbell Hall, University of California, Berkeley, CA 94720 \\ 
  \email{areiners@astron.berkeley.edu} 
  \and
  Hamburger Sternwarte, Universit\"at Hamburg, Gojenbergsweg 112, 21029 Hamburg, Germany \\
}
\date{\today} 

\titlerunning{Differential rotation with temperature and rotation}

\abstract{More than 600 high resolution spectra of stars with spectral
  type F and later were obtained in order to search for signatures of
  differential rotation in line profiles. In 147 stars the rotation
  law could be measured, with 28 of them found to be differentially
  rotating. Comparison to rotation laws in stars of spectral type A
  reveals that differential rotation sets in at the convection
  boundary in the HR-diagram; no star that is significantly hotter
  than the convection boundary exhibits the signatures of differential
  rotation.  Four late A-/early F-type stars close to the convection
  boundary and at $v\,\sin{i} \approx 100$\,km\,s$^{-1}$ show
  extraordinarily strong absolute shear at short rotation periods
  around one day.  It is suggested that this is due to their small
  convection zone depth and that it is connected to a narrow range in
  surface velocity; the four stars are very similar in $T_{\rm eff}$
  and $v\,\sin{i}$. Detection frequencies of differential rotation
  $\alpha = \Delta\Omega/\Omega > 0$ were analyzed in stars with
  varying temperature and rotation velocity. Measurable differential
  rotation is more frequent in late-type stars and slow rotators. The
  strength of absolute shear, $\Delta\Omega$, and differential
  rotation $\alpha$ are examined as functions of the stellar effective
  temperature and rotation period. The highest values of
  $\Delta\Omega$ are found at rotation periods between two and three
  days. In slower rotators, the strongest absolute shear at a given
  rotation rate $\Delta\Omega_{\rm max}$ is given approximately by
  $\Delta\Omega_{\rm max} \propto P^{-1}$, i.e., $\alpha_{\rm max}
  \approx$\,const. In faster rotators, both $\alpha_{\rm max}$ and
  $\Delta\Omega_{\rm max}$ diminish less rapidly. A comparison with
  differential rotation measurements in stars of later spectral type
  shows that F-stars exhibit stronger shear than cooler stars do and
  the upper boundary in absolute shear $\Delta\Omega$ with temperature
  is consistent with the temperature-scaling law found in Doppler
  Imaging measurements.
  
}

\maketitle
                                %


\section{Introduction}
\label{sect:introduction}

Stellar rotation rates range from those too slow to be detected by
Doppler broadening up to rates at which centrifugal forces become
comparable to surface gravity.  Surface magnetic fields, either fossil
or generated by some type of magnetic dynamo, can couple to ionized
plasma and brake a star's rotation via a magneto-thermal wind.
Magnetic braking is observed in stars with deep convective envelopes
where magnetic dynamo processes can efficiently maintain strong
magnetic fields. As a consequence, field stars of a spectral type
later than F are generally slow rotators with surface velocities below
10\,km\,s$^{-1}$.

Magnetic braking requires the existence of a magnetic field, which
also causes the plethora of all the effects found in stellar magnetic
activity.  While early magnetic braking may be due to fossil fields
amplified during contraction of the protostellar cloud, magnetic
activity at later phases requires a mechanism that maintains magnetic
fields on longer timescales. In the Sun, a magnetic dynamo located at
the interface between convective envelope and radiative core is driven
by radial differential rotation. This dynamo has been identified as
the main source of magnetic fields, although there is growing evidence
that it is not the only source of magnetic field generation
\citep[e.g.,][]{Schrijver00} and that magnetic fields also exist in
fully convective stars \citep{Johns-Krull96}. The solar-type dynamo,
however, has the potential to generate magnetic fields in all
non-fully convective stars, as long as they show a convective
envelope, and magnetic fields have been observed in a variety of
slowly rotating stars \citep[e.g.,][]{Marcy84, Gray84, Solanki91}.

Calculations of stellar rotation laws, which describe angular velocity
as a function of radius and latitude, have been carried out by
\cite{Kitchatinov99} and \cite{Kueker05} for different equatorial
angular velocities. \cite{Kitchatinov99} investigated rotation laws in
a G2- and a K5-dwarf. They expect stronger latitudinal differential
rotation in slower rotators, and their G2-dwarf model exhibits
stronger differential rotation than the K5-dwarf does.
\cite{Kueker05} calculate a solar-like model, as well as a model of an
F8 main sequence star. They also come to the conclusion that
differential rotation is stronger in the hotter model; the maximum
differential rotation in the F8 star is roughly twice as strong as in
the G2 star for the same viscosity parameter. The calculated
dependence of horizontal shear on rotation rate, however, does not
show a monotonic slope but has a maximum that occurs near 10\,d in the
F8 type star and around 25\,d in the solar-type star. The strength of
differential rotation depends on the choice of the viscosity
parameter, which is not well constrained, but the trends in
temperature and rotation are unaffected by that choice.

Observational confirmation of solar radial and latitudinal
differential rotation comes from helioseismological studies that
provide a detailed picture of the differentially rotating outer
convection zone \citep[e.g.,][]{Schou98}. Such seismological studies
are not yet available for any other star. Asteroseismological
missions, like COROT and Kepler, may open a new window on stellar
differential rotation, but its data quality may provide only a very
limited picture in the near future \citep{Gizon04}.  In the case of
the Sun, radial differential rotation manifests itself in latitudinal
differential rotation that can be observed at the stellar surface, but
all stars except the Sun are at distances where their surfaces cannot
be adequately resolved.  With the advent of large optical
interferometers, more may be learned from observations of spatially
resolved stellar surfaces \citep{Domiciano04}. For now, we have to
rely on indirect methods to measure the stellar rotation law.

Photometric programs that search for stellar differential rotation
assume that starspots emerge at various latitudes with different
rotation rates, as observed on the Sun. \cite{Hall91} and
\cite{Donahue96} measured photometric rotation periods and interpreted
seasonal variations as the effect of differential rotation on
migrating spots.  Although these techniques are comparable to the
successful measurements of solar differential rotation through
sunspots \citep{Balthasar86}, they still rely on a number of
assumptions, e.g., the spot lifetime being longer than the
observational sequence, an assumption difficult to test in stars other
than the Sun \cite[cf.][]{Wolter05}. Photometric measurements report
lower limits for differential rotation on the order of 10\% of the
rotation velocity (i.e. the equator rotating 10\% faster than the
polar regions), and differential rotation is reported to be stronger
in slower rotators.

Doppler Imaging (DI) has been extensively used to determine
latitudinal differential rotation where the derived maps are
constructed from time-series of high-resolution spectra. Differential
rotation can then be extracted from comparing two surface maps taken
with time separation of a few rotation periods
\cite[e.g.,][]{Donati97,Wolter05}. Doppler maps can also be
constructed that incorporate differential rotation during the
inversion algorithm \citep{Petit02}. Detections of differential
rotation through DI have recently been compiled by \cite{Barnes05},
who also analyze dependence on stellar rotation and temperature, and
then compare them to results obtained from other techniques. Their
results will be discussed in Sect.\,\ref{sect:results}.

The technique employed in this paper is to search for latitudinal
differential rotation in the shape of stellar absorption line
profiles. This method is applicable only to stars not dominated by
spots. From a single exposure, latitudinal solar-like differential
rotation -- i.e. the equator rotating faster than polar regions -- can
be derived by measuring its unambiguous fingerprints in the Fourier
domain.  The foundations of the Fourier transform method (FTM) were
laid by \cite{Gray77,Gray98}, and first models were done by
\cite{Bruning81}. A detailed description of the fingerprints of
solar-like differential rotation and the first successful detections
are given in \cite{Reiners02a, Reiners03a}. The FTM is limited to
moderately rapid rotators (see Sect.\,\ref{sect:method}), but the big
advantage of this method is that latitudinal differential rotation can
be measured from a single exposure. This allows the analysis of a
large sample of stars with a comparably small amount of telescope
time.  \cite{Reiners03a} report on differential rotation in ten out of
a sample of 32 stars of spectral types F0--G0 with projected rotation
velocities $12$\,km\,s$^{-1} < v\,\sin{i} < 50$\,km\,s$^{-1}$.
\cite{Reiners03b} investigated a sample of 70 rapid rotators with
$v\,\sin{i} > 45$\,km\,s$^{-1}$ and found a much lower fraction of
differential rotators.  Differential rotation has also been sought in
A-stars that have no deep convective envelopes.  \cite{Reiners04}
report on three objects out of 76 in the range A0--F1,
$60$\,km\,s$^{-1} < v\,\sin{i} < 150$\,km\,s$^{-1}$, which show
signatures of differential rotation.

In the cited works, differential rotation is investigated in stars of
limited spectral types and rotation velocities. In this paper, I aim
to investigate all measurements of differential rotation from FTM, add
new observations, and finally compare them to results from DI and
theoretical predictions. Currently, more than 600 stars were observed
during the course of this project, and in 147 of them the rotation law
could be measured successfully.

\section{Differential rotation in line profiles}
\label{sect:method}

Latitudinal differential rotation has a characteristic fingerprint in
the shape of the rotational broadening that appears in each spectral
line.  Since all other line broadening mechanisms, like turbulence,
thermal, and pressure broadening, etc., also affect the shape of
spectral lines, the effects of differential rotation are very subtle.
Signatures of differential rotation and a recipe for measuring them
are presented in \cite{Reiners02a}. Since the signatures are so small,
and spectral line blending is a serious issue even in stars of
moderate rotation rates at $v\,\sin{i} \approx 20$\,km\,s$^{-1}$,
single lines are not measured, but instead a total broadening function
is constructed from many lines of similar intrinsic shape.  This
process typically involves 15 lines in slow rotators observed at very
high resolution (Sect.\,\ref{sect:observations}), and 300 lines in
rapid rotators observed at lower resolution.  In order to derive a
unique broadening function at the required precision, high data
quality is required.  Detailed information about demands on data
quality can be found in \cite{Reiners03a, Reiners03b}.

Interpretation of the profile's shape with rotational broadening
requires that the line profiles are not affected by starspots, stellar
winds, spectroscopic multiplicity, etc.. The Fourier transform method
is therefore limited to unspotted single stars with projected rotation
velocities $v\,\sin{i} \ga 12$\,km\,s$^{-1}$, and an upper limit at
$v_{\rm eq} \approx 200$\,km\,s$^{-1}$ is set by gravitational
darkening. More specific information on the influence of starspots, of
very rapid rotation, and some examples of detected signatures of
differential rotation in line profiles can be found in
\cite{Reiners03}, and \cite{Reiners02a, Reiners03a, Reiners03b}.

From the derived broadening profile, the rotation law is determined by
measuring the first two zeros of the profile's Fourier transform,
$q_1$ and $q_2$. In sufficiently rapid rotators, those are direct
indicators of rotational broadening since other broadening mechanisms
(for example turbulence or instrumental broadening) do not show zeros
at such low frequencies.  This is also the reason only stars with
$v\,\sin{i} \ga 12$\,km\,s$^{-1}$ can be studied; in slower rotators,
line broadening is dominated by turbulence and the zeros due to
rotation cannot be measured. In stars spinning fast enough to be
analyzed with FTM, approximation of net broadening by convolutions is
also justified. The important point in choosing the Fourier domain for
profile analysis is that convolutions become multiplications in
Fourier space. Thus, the fingerprints of \emph{rotational} broadening
are directly visible in the observed broadening profile's Fourier
transform, and the spectra do not have to be corrected for
instrumental or for any other line broadening, as long as the targets'
rotation dominates the important frequency range.

In the following, the stellar rotation law will be approximated in
analogy to the solar case. Differential rotation is expressed in terms
of the variable $\alpha$, with $\Omega$ the angular velocity and $l$
the latitude. The rotation law is approximated by
\begin{eqnarray}
  \label{eq:diff}
  \Omega(l) & = & \Omega_{Equator} \cdot \alpha \sin^{2}(l),\\
  \alpha & = & \frac{\Omega_{Equator} - \Omega_{Pole}}{\Omega_{Equator}} = \frac{\Delta\Omega}{\Omega}.
\end{eqnarray}

Solar-like differential rotation is characterized by $\alpha > 0$
$(\alpha_\odot \approx 0.2)$. Sometimes $\Delta \Omega$ is called
\emph{absolute} differential rotation. To avoid confusion with
\emph{relative} differential rotation $\alpha = \Delta \Omega/\Omega$,
I will refer to $\Delta \Omega$ as absolute shear and to $\alpha$ as
differential rotation. $\Delta \Omega$ is given in units of
rad\,d$^{-1}$. As shown in \cite{Reiners03a}, the parameter
$\alpha/\sqrt{\sin{i}}$, with $i$ the inclination of the stellar
rotation axis, can be directly obtained from the ratio of the Fourier
transform's first two zeros, $q_2/q_1$.  Thus, determination of the
stellar rotation law in Eq.\,\ref{eq:diff} from a single stellar
spectrum is a straightforward exercise.

\subsection{Variables to describe differential rotation}

The role of differential rotation especially for magnetic field
generation in rapid rotators is not understood well. In the case of
the Sun, we know that its magnetic cycle is driven, at least in part,
by radial differential rotation, which itself is reflected in
latitudinal differential rotation. How (and if) magnetic dynamos
depend on the strength of differential rotation in terms of $\alpha$
or on absolute shear $\Delta\Omega$ has not yet been empirically
tested.  Furthermore, different observing techniques measure different
quantities, and authors express their results on stellar rotation laws
in different variables. A variable frequently used is the lap time,
i.e. the time it takes the equator to lap the polar regions (or vice
versa), which is essentially the reciprocal of the shear
$\Delta\Omega$.  Differential rotation $\alpha=\Delta\Omega/\Omega$ is
the shear divided by the angular velocity. Expressing rotation
velocity in terms of rotation period $P$ essentially means the
reciprocal of angular velocity $\Omega$, but using $\Delta P$ instead
of $\Delta \Omega$ introduces several problems, since one has to
consider whether $P$ denotes equatorial or polar rotation period, the
latter being larger with solar-like differential rotation
\citep[cp.][]{Reiners03b}. In this paper, I will express the rotation
law in terms of $\alpha$ and $\Delta\Omega$, in order to search for
correlations with rotation velocity. The quantity measured by the FTM
is $\alpha$ with a fixed observational threshold of $\alpha_{\rm min}
\approx 0.05$ thereby limiting the detection of deviations to rigid
rotation. The uncertainty in $\alpha$ measured by FTM is approximately
$\delta\alpha = 0.1$. The perhaps more intuitive parameter for the
physical consequences of differential rotation, however, is the shear
$\Delta\Omega = \alpha\Omega$, where $\Omega$ has to be obtained from
$v\,\sin{i}$ and the radius.  When analyzing the rotation law in terms
of $\Delta\Omega$, one has to keep in mind that it can never exceed
angular velocity itself, if polar and equatorial regions are not
allowed to rotate in opposite directions.  In other words,
differential rotation cannot be larger than 100\%.

\section{Observational data}
\label{sect:observations}

\begin{table*}
  \caption{\label{tab:observations}Data used for this analysis. Spectra
    were taken with different high-resolution spectrographs.
    Detection of differential rotation requires higher resolution in
    slower rotators. The minimum rotation velocities required for
    determining the rotation law are given in column 4. From the
    147 spectra from which differential rotation could be determined, 28
    stars show signatures of differential rotation} 
  \centering
  \begin{tabular}{crcccc}
    \hline
    \hline
    \noalign{\smallskip}
    Instrument & Resolving power $\lambda/\Delta\lambda$ & Wavelength Range [\AA]& min. $v\,\sin{i}$ [km/s] & total \# of stars & \# of differential rotators\\
    \noalign{\smallskip}
    \hline
    \noalign{\smallskip}
    CES & 220\,000 & 40 & 12 & 32 & 14\\
    FEROS & 48\,000 & 6000 & 45 & 23 & 3\\
    FOCES & 40\,000 & 6000 & 45 & 81 & 7\\

    FLAMES/UVES  & 47\,000 & 2000 & 45 & 11 & 4\\
    \noalign{\smallskip}
    \hline
    \noalign{\smallskip}
    total &&&& 147 & 28\\
    \noalign{\smallskip}
    \hline
  \end{tabular}
\end{table*}

Data for this study have been compiled from observations carried out
at different telescopes. Observations of field stars with projected
rotation velocities higher than $v\,\sin{i} = 45$\,km\,s$^{-1}$ have
been carried out with FEROS on the 1.5m telescope ($R = 48\,000$),
ESO, La~Silla, or with FOCES ($R = 40\,000$) on the 2.2m at CAHA.
Slower rotating field stars were observed at higher resolution at the
CES with the 3.6m telescope, ESO, La~Silla ($R = 220\,000$).
Additionally, observations in open clusters were carried out with the
multi-object facility FLAMES feeding VLT's optical high-resolution
spectrograph UVES at a resolution of $R = 47\,000$.  Details of the
spectra and instruments used are given in
Table\,\ref{tab:observations}. Parts of the data were published in
\cite{Reiners03a, Reiners03b} and \cite{Reiners05}. For more
information about observations and data reduction, the reader is
referred to these papers. Ten of the cluster targets (FLAMES/UVES), as
well as 34 FOCES, targets were not reported on in former
publications\footnote{It should be noted that the FLAMES/UVES targets
  observed in open cluster fields are not necessarily cluster members,
  as was shown for example for Cl*~IC\,4665~V\,102 in
  \cite{Reiners05}. A detailed investigation of cluster membership
  goes beyond the scope of this analysis; rotation velocity and
  spectral type make them sufficiently comparable to the (probable)
  field stars in this context.}.

More than 600 stars were observed for this project during the last
four years. For this analysis, I selected the 147 stars for which the
ratio of the Fourier transform's zeros $q_2/q_1$ -- the tracer of the
rotation law -- is measured with a precision better than 0.1, i.e.
better than 6\% in case of the typical value of $q_2/q_1 = 1.76$.
These 147 stars exhibit broadening functions that are (i) symmetric
(to avoid contamination by starspots), and (ii) reveal rotation
velocities between $v\,\sin{i} = 12$\,km\,s$^{-1}$ and
150\,km\,s$^{-1}$, the latter being an arbitrary threshold in order to
minimize the amount of gravity-darkened stars with $v_{\rm eq} \gg
200$\,km\,s$^{-1}$ \citep{Reiners03}.  Depending on the spectral
resolution, the minimum rotation velocity is higher than $v\,\sin{i} =
12$\,km\,s$^{-1}$ (CES) or 45\,km\,s$^{-1}$ (FEROS, FOCES, UVES, see
also Table\,\ref{tab:observations}).

More than 450 stars have not been used in this analysis. Their
broadening functions show a whole variety of broadening profiles. Many
are slow rotators, although some were reported as rapid rotators in
earlier catalogues \citep[cf.][]{Reiners03b}.  A large number of
spectroscopic binaries (or even triples) were found, and another large
part of the sample shows spectra that are apparently distorted by
starspots or other mechanisms that cause the spectra to appear
asymmetric. For analyzing the rotation law from the broadening profile
alone, these spectra are useless and are not considered in this work.
They may be promising targets for DI techniques or other science; a
catalogue of broadening profiles is in preparation.

\cite{Reiners04} measured the rotation law in 78 stars of spectral
type A including a few early F-type stars. This sample will only be
incorporated in the HR-diagram in Sect.\,\ref{sect:HRD}. As will be
shown there, rotation laws in A-stars are fundamentally different from
those in F-stars. For this reason, I will not incorporate the A-stars
in the sample analyzed for rotation- and temperature-dependencies of
differential rotation. One of the stars observed during this project
has recently been studied with DI, with comparison of the results
given in the Appendix in Sect.\,\ref{sect:DI}.

\section{Accuracy of rotation law measurements}

Projected rotation velocities $v\,\sin{i}$ were derived from the first
zero $q_1$ in the Fourier transform as explained in \cite{Reiners03b}.
As mentioned there, the precision of this measurement is usually $<
1$\,km\,s$^{-1}$, but simulations revealed a systematic uncertainty of
$\sim 5\%$ in $v\,\sin{i}$. Thus, I chose the maximum of the
uncertainty in the intrinsic measurement and the $5\%$ limit as the
error on $v\,\sin{i}$.

The typical uncertainty for the determined values of relative
differential rotation in terms of $\alpha = \Delta\Omega/\Omega$ is
$\delta\alpha \approx 0.1$ \citep{Reiners03a}. In slow rotators, this
uncertainty is dominated by the noise level due to comparably sparse
sampling of the profile even at very high resolution. In fast
rotators, the profile (i.e., the zeros of the profile's Fourier
transform) can be measured with very high precision.  Here the
uncertainty in $\alpha$ stems from poor knowledge of the limb
darkening parameter $\epsilon$. For the case of a linear limb
darkening law, the value of the measured ratio of the Fourier
transform's first two zeros, $q_1$ and $q_2$, is in the range $1.72 <
q_2/q_1 < 1.85$ \citep{Dravins90}. With the very conservative estimate
that $\epsilon$ is essentially unknown ($0.0 < \epsilon < 1.0$), every
star with $1.72 < q_2/q_1 < 1.85$ in this analysis was considered a
rigid rotator, and $q_2/q_1 < 1.72$ was interpreted as solar-like
differential rotation with the equator rotating faster than the polar
regions. From the measured value of $q_2/q_1$, the parameter
$\alpha/\sqrt{\sin{i}}$ was determined as explained above.

Stars can exhibit ratios of $q_2/q_1 > 1.85$ as well, but this value
also does not agree with rigid rotation of a homogenous stellar
surface. In the investigated sample of 147 stars, eight (5\%) exhibit
a ratio of $q_2/q_1 > 1.85$. In contrast to solar-like differential
rotation with the equator rotating more rapidly than polar regions,
this case may be due to anti solar-like differential rotation with
polar regions rotating faster than the equator. On the other hand, it
can also be caused by a cool polar cap, which is expected to occur in
rapidly rotating stars \citep{Schrijver01}. The lower flux emerging
from a cool polar cap makes the line center shallow and has the same
signature as anti solar-like differential rotation. While anti
solar-like differential rotation cannot be distinguished from a cool
polar cap by investigating the rotation profile, the existence of a
cool polar cap in rapidly rotating F-stars seems much more plausible
than anti solar-like differential rotation. Differentiation between
the two cases can only be achieved by measuring differential rotation
independent of the line shape. For the scope of this paper, however, I
will interpret stars with $q_2/q_1 > 1.85$ as rigid rotators with a
cool polar cap.

\section{Rotation velocity and spectral type}
\label{sect:rotation-color}

\begin{table}
  \caption{\label{tab:vs_BV}Amount of differential rotators, total number of stars, and percentage of differential rotators for each subsample of Fig.\,\ref{plot:vsini_BV}}
  \centering
  \begin{tabular}{crrr}
    \hline
    \hline
    \noalign{\smallskip}
    Subsample & \# of diff. rotators & total \# of stars & percentage\\
    \noalign{\smallskip}
    \hline
    \noalign{\smallskip}
    I   & 2 & 57 & 4\%\\
    II  & 4 & 24 & 17\%\\
    III & 5 & 21 & 24\%\\
    IV  & 17 & 45 & 38\%\\
    \noalign{\smallskip}
    \hline
  \end{tabular}
\end{table}
\begin{figure}
  \centering
  \resizebox{\hsize}{!}{\includegraphics[angle=-90]{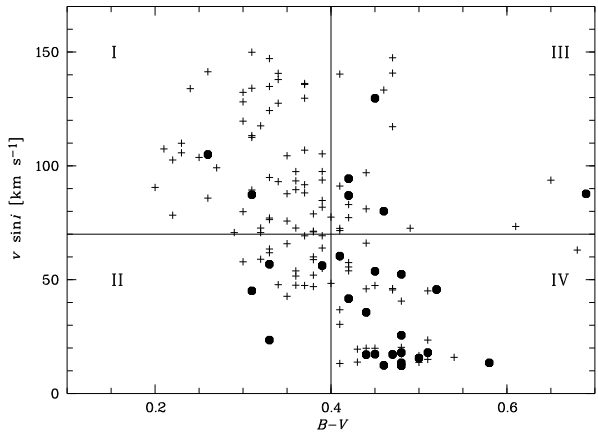}}
  \caption{\label{plot:vsini_BV}Projected rotation velocity plotted vs. $B-V$ for
    the whole sample of 147 stars. Differential rotators are plotted
    as filled circles. Contents of subsamples I--IV are given in
    Table\,\ref{tab:vs_BV}. }
\end{figure}

The observations carried out with FEROS, FOCES, UVES, and the CES are
very homogeneous in terms of data quality and the investigated
wavelength region, and only measurements that fulfill the criterion
$\delta\alpha < 0.1$ are considered.  The targets have similar
spectral types, but they do not form a statistically unbiased sample.
The quality requirements discussed above make it difficult to analyze
stars in a well-defined sample. This sample certainly is severely
biased by observational and systematic effects. The most important
bias is probably due to rotational braking, by connecting spectral
type with rotational velocity in the presumably old field stars. Later
spectral types suffer from stronger magnetic braking and are expected
to be generally slower than earlier spectral types.

Projected rotation velocities $v\,\sin{i}$ of the sample stars are
plotted versus color $B-V$ in Fig.\,\ref{plot:vsini_BV}, while
differentially rotating stars (i.e. stars with $\alpha > 0$) are
indicated by filled circles and will be discussed in detail in
Sect.\,\ref{sect:results}. A clear dependence of $v\,\sin{i}$ on color
is apparent, as expected. In Fig.\,\ref{plot:vsini_BV}, the sample is
divided arbitrarily into four subsamples with projected rotation
velocities that are higher (lower) than $v\,\sin{i} =
70$\,km\,s$^{-1}$, and color redder (bluer) than $B-V = 0.4$. The
total numbers of stars in each subsample and the numbers of
differential rotators are given in Table\,\ref{tab:vs_BV}. Most
targets occupy regions I and IV. The scarcity in region II is due to
the early F-type stars not being subject to strong magnetic braking;
since the sample mainly consists of field stars, most stars later than
$B-V = 0.4$ have been decelerated into region IV and do not appear in
region III. Thus, slower rotating stars generally have a later
spectral type in the sample. This implies that the effects of
temperature and rotation velocity on the fraction of differentially
rotating stars are degenerate in this sample; slow rotation implies
late spectral type. It is therefore not possible to uniquely
distinguish between the effects of rotation and spectral type in the
investigated sample. This degeneracy has to be kept in mind when
trying to interpret rotation- and temperature-dependencies of
differential rotation in the following chapters.

\section{Results}
\label{sect:results}

\begin{figure*}
  \centering 
  \resizebox{\hsize}{!}{\includegraphics[angle=-90]{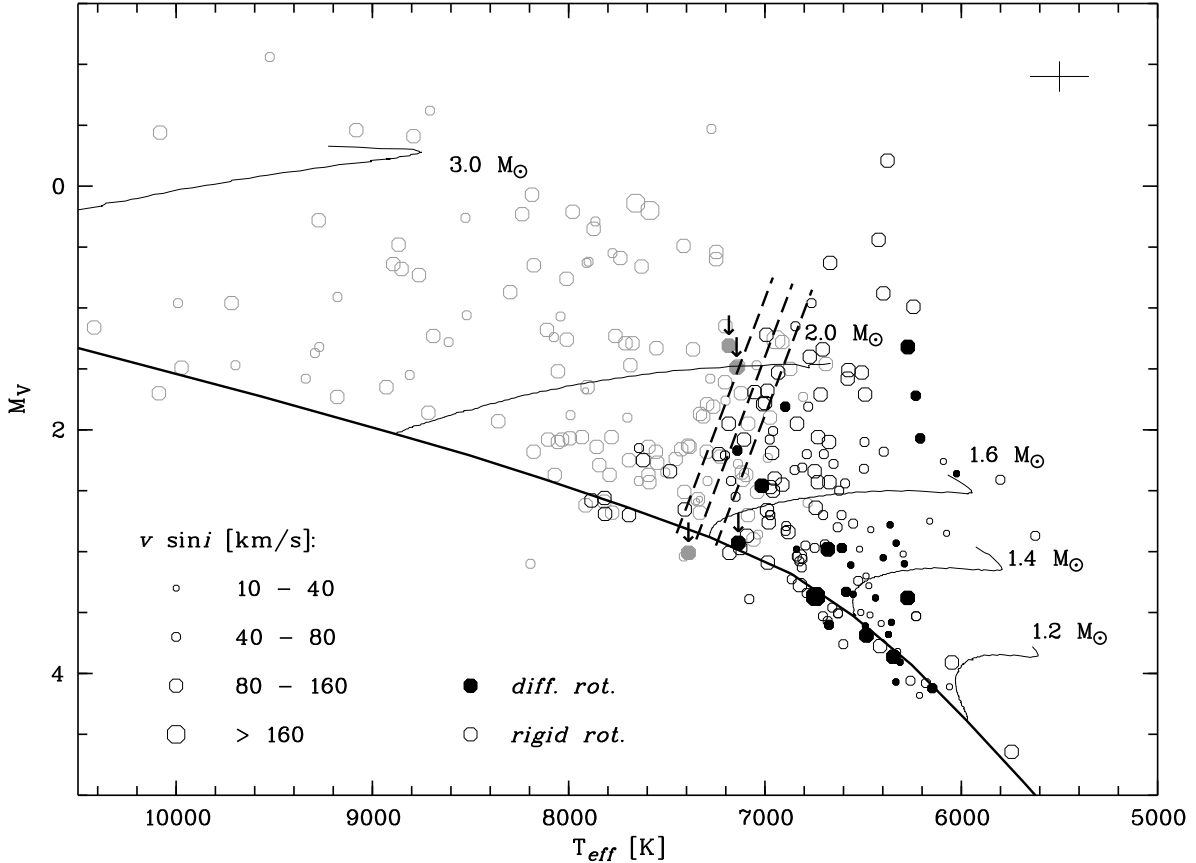}}
  \caption{\label{plot:HRD}HR-diagram of all currently available measurements
    of latitudinal differential rotation from FTM. Stars analyzed in
    this work are plotted in black, and stars from the A-star sample
    in \cite{Reiners04} in grey. Differential rotators are indicated
    by filled circles. Symbol size represents ranges of projected
    rotation velocity $v\,\sin{i}$ as explained in the figure. A
    typical error bar is given in the upper right corner.
    Evolutionary tracks and the ZAMS from \cite{Siess00} are
    overplotted. Dashed lines mark the area of the granulation
    boundary according to \cite{Gray89}, while no deep convection
    zones are expected on the hot side of the boundary. Downward
    arrows indicate the four stars from Table\,\ref{tab:A-stars} (see
    text).}
\end{figure*}

As a first result of the profile analysis, I present $v\,\sin{i}$ and
the measurement of the rotation law for all 147 objects in
Appendix\,\ref{sect:appendix}. Fields stars (FEROS, FOCES, and CES
observations) are compiled in Table\,\ref{tab:fieldstars}, FLAMES/UVES
targets in Table\,\ref{tab:uvesstars}.

In the sample of 147 stars for which the rotation law was measured, 28
(19\%) exhibit signatures of solar-like differential rotation
($q_2/q_1 < 1.72$). CES-, FEROS-, and FOCES-samples contain field
stars of spectral type later than F0, with the majority of them
brighter than $V = 6$\,mag. Stars in fields of open clusters contained
in the UVES-sample probably are mostly younger than the field stars,
and all of them should have reached the main sequence
\citep{Stahler04}.

\subsection{Differential rotation in the HR-diagram}
\label{sect:HRD}

All currently available measurements of the stellar rotation law from
rotation profile shape \citep[i.e, 147 stars from this sample and 78
stars from ][]{Reiners04} are plotted in an HR-diagram in
Fig.\,\ref{plot:HRD}. For the field and A-stars, effective temperature
$T_{\rm eff}$ and bolometric magnitude M$_{\rm bol}$ were derived from
$uvby\beta$ photometry taken from \citet{Hauck98} using the program
UVBYBETA published by \citet{Moon85a}. For $T_{\rm eff}$ a new
calibration by \citet{Napiwotzki93} based on the grids of
\citet{Moon85b} was used, and the statistical error of the temperature
determination is about $\Delta T_{\rm eff} = 150$\,K. For three of the
field stars, no $uvby\beta$ photometry is available; all three are
rigid rotators, so no value of $T_{\rm eff}$ was calculated to avoid
inconsistencies. For the cluster stars, no $uvby\beta$ data is
available, so radius and temperature are estimated from $B-V$ color
using zero-age main sequence (ZAMS) polynomials taken from
\cite{Gray76}, i.e., they are assumed to be young. For them, M$_{\rm
  bol}$ is calculated from
\begin{equation}
  \label{eq:rmbol}
  {\rm M}_{\rm bol} = 42.36 - 10 \log{T_{\rm eff}} - 5 \log{R/R_{\odot}},
\end{equation}
with $R_{\odot}$ the solar radius. To get absolute V-magnitudes M$_V$
from bolometric magnitudes M$_{\rm bol}$, the bolometric correction
calibrated from \cite{Hayes78} was applied.

In Fig.\,\ref{plot:HRD}, stars are plotted as open or filled circles
that indicate rigid ($\alpha = 0$) or differential rotation ($\alpha >
0$), respectively. Circle sizes display projected rotation velocities
$v\,\sin{i}$ as explained in the figure.  The 147 stars given in
Tables\,\ref{tab:fieldstars} and \ref{tab:uvesstars} are plotted as
black symbols, while stars from \cite{Reiners04} are plotted as grey
symbols.  Evolutionary main sequence tracks for $M = 1.2, 1.4, 1.6,
2.0$, and $3.0\,$M$_\odot$ and the ZAMS from \cite{Siess00} are also
shown.  Near the interface of A-stars and F-stars, the ``granulation
boundary'' from \cite{Gray89} is indicated with dashed lines. This is
the region where line bisectors measured in slow rotators show a
reversal. For dwarfs and subgiants, the ``granulation boundary''
coincides with theoretical calculations of the ``convection boundary''
and thus can be identified as the region where deep envelope
convection disappears \citep[cf.][]{Gray89}. The exact region of this
boundary is not defined well and may depend on factors other than
temperature and luminosity.

The stars shown in Fig.\,\ref{plot:HRD} cover a wide range in
temperature on both sides of the convection boundary. Because field
stars are investigated, many stars have evolved away from the ZAMS.
Rotation velocity as indicated by symbol size follows the well-known
behavior of magnetic braking; late type stars are generally slowed
down and exhibit slower rotation velocities. The striking fact visible
in Fig.\,\ref{plot:HRD} is that none of the differential rotators
detected with FTM lies significantly on the hot side of the convection
boundary -- latitudinal differential rotation has only been detected
in stars believed to possess a deep convective envelope.

The largest group of differential rotators can be found near the ZAMS
at all temperatures that are cooler than the convection boundary. The
convection boundary itself is also populated by differential rotators.
A few others can be found far away from the ZAMS (at masses $\ga
1.6$\,M$_\odot$ ) to the right end of the main sequence tracks.  There
is a hint that differential rotators are lacking in the region between
these few rotators and those near the ZAMS and again between them and
those near the convection boundary, although this region is occupied
by rigidly rotating stars (i.e., stars with differential rotation
weaker than the observational threshold). However, it is not clear
from the available sample whether temperature (and evolutionary stage)
or rotation velocity may be the important parameter in determining
differential rotation in F-type stars (see
Sect.\,\ref{sect:rotation-color}).

\subsection{Differentially rotating A-stars at the convection boundary}
\label{sect:Astars}

\begin{table}
  \caption{\label{tab:A-stars}Four stars with extreme values of differential 
    rotation $\Delta\Omega$. It is speculated that these objects represent a special class of targets (marked with downward arrows in Fig.\,\ref{plot:HRD}) in a narrow region of rotation velocity and effective temperatures near the convection boundary}
  \begin{tabular}{cccccc}
    \hline
    \hline
    \noalign{\smallskip}
    Star & Sp & $v\,\sin{i}$ & $\Delta\Omega \sin{i}$ & $T_{\rm eff}$ & M$_V$\\
    & & [km\,s\,$^{-1}$] & [rad\,d$^{-1}$] & [K]\\
    \noalign{\smallskip}
    \hline
    \noalign{\smallskip}
    HD~\quad\phantom{00}6869 & A9 & $\phantom{0}95\pm\phantom{0}5^1$ & $2.3\pm1.5^1$ & 7390 & 3.01\\
    HD~\quad\phantom{0}60555 & A6 & $109\pm\phantom{0}5^1$ & $1.3\pm0.7^1$ & 7145 & 1.49\\
    HD~\quad109238 & F0 & $103\pm\phantom{0}4^1$ & $1.3\pm0.6^1$ & 7184 &  1.31\\
    IC\,4665~V\,102 & A9 & $105\pm12^2$ & $3.6\pm0.8^2$ & 7136 & 2.93\\
    \noalign{\smallskip}
    \hline
  \end{tabular}
  \begin{list}{}{}
  \item[$^1$]\cite{Reiners04}
  \item[$^2$]\cite{Reiners05}
  \end{list}
\end{table}

Among all the differentially rotating stars shown as filled circles in
Fig.\,\ref{plot:HRD}, the four stars with the strongest shear
$\Delta\Omega$ are found very close to the convection boundary.  This
group of rapidly rotating late A-/early F-type stars is listed in
Table\,\ref{tab:A-stars}\footnote{It should be noted that the only
  star in this group, that lies on the cool side of the convection
  boundary, is Cl*~IC\,4665~V\,102. Temperature and luminosity were
  estimated from $B-V$ color using ZAMS polynomials.
  Cl*~IC\,4665~V\,102, however, is not a member of the cluster
  IC\,4665 \citep{Reiners05}, hence its position in the HR-diagram is
  most uncertain. It may have a much higher absolute luminosity that
  shifts it even closer towards the convection boundary.}. All four
exhibit very similar effective temperatures around $T_{\rm eff} =
7200$\,K, putting them into the region where convective envelopes are
extremely shallow.  The four stars show remarkably similar projected
rotation velocities all within 10\% of $v\,\sin{i} =
100$\,km\,s$^{-1}$, and all four have rotation periods shorter than $P
= 2$\,d.  Two of them, HD\,6869 and Cl*~IC\,4665~V\,102, exhibit a
shear that is as strong as $\Delta\Omega > 2$\,rad\,d$^{-1}$ at
rotation periods that are shorter than one day. This contrasts the 34
other stars that show comparably small rotation periods but have
different temperatures or surface velocities. None of the 34 other
rapid rotators exhibits a shear $\Delta\Omega\sin{i} $ in excess of
$0.7$\,rad\,d$^{-1}$, i.e. a factor of three weaker than the two
mentioned above, and only three of the 34 show $\Delta\Omega > 0$ at
all.

The fact that the four strongest differential rotators are found at
hot temperatures and high rotation velocities contradicts the general
trend that differential rotation is more common in slowly rotating
cool stars, which will be discussed in the following sections. All
four are very close to the convection boundary or even on its blue
side, meaning extremely shallow convective envelopes.  This leads to
the suspicion that the mechanism responsible for the strong shear is
different from the one that drives the shear in stars with deeper
convection zones. This is supported by the observation that all four
stars also exhibit very similar (and comparably large) surface
velocities of about 100\,km\,s$^{-1}$. It is thus suggested that the
strong surface shear in the four stars in Table\,\ref{tab:A-stars} is
not comparable to the rest of the sample, but is facilitated by the
high surface velocity in a particularly shallow convective envelope.
Such a mechanism could be supported by eigenmodes that are comparable
to pulsational instabilities, for example in $\delta$~Scu stars, but
further investigation is beyond the scope of this paper.

\subsection{The fractional amount of differential rotators}

\begin{figure*}
  \centering \mbox{
    \resizebox{.45\hsize}{!}{\includegraphics[angle=-90,bbllx=28,bblly=50,bburx=270,bbury=650]{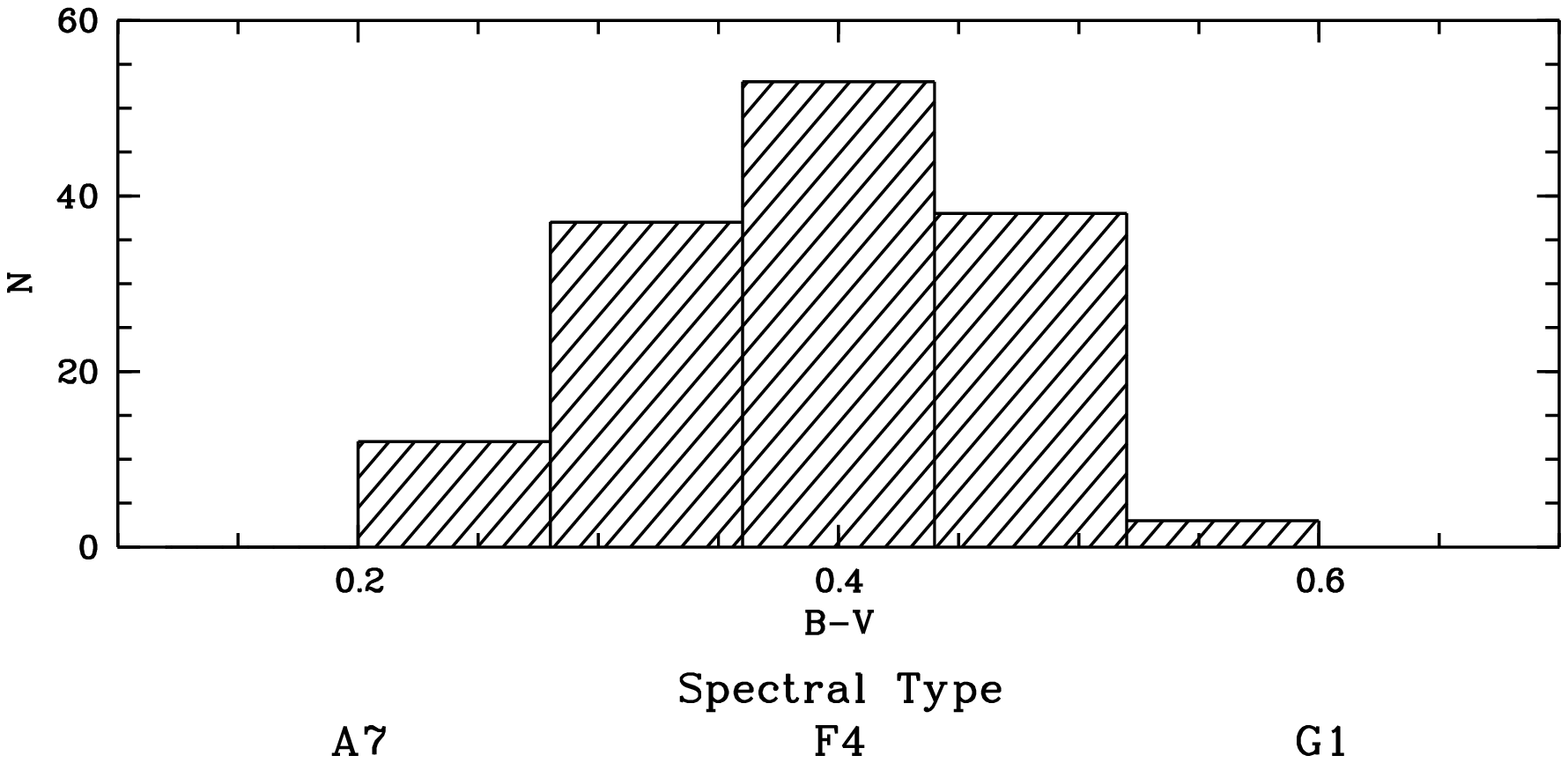}}\hspace{0.05\hsize}
    \resizebox{.45\hsize}{!}{\includegraphics[angle=-90,bbllx=28,bblly=50,bburx=270,bbury=650]{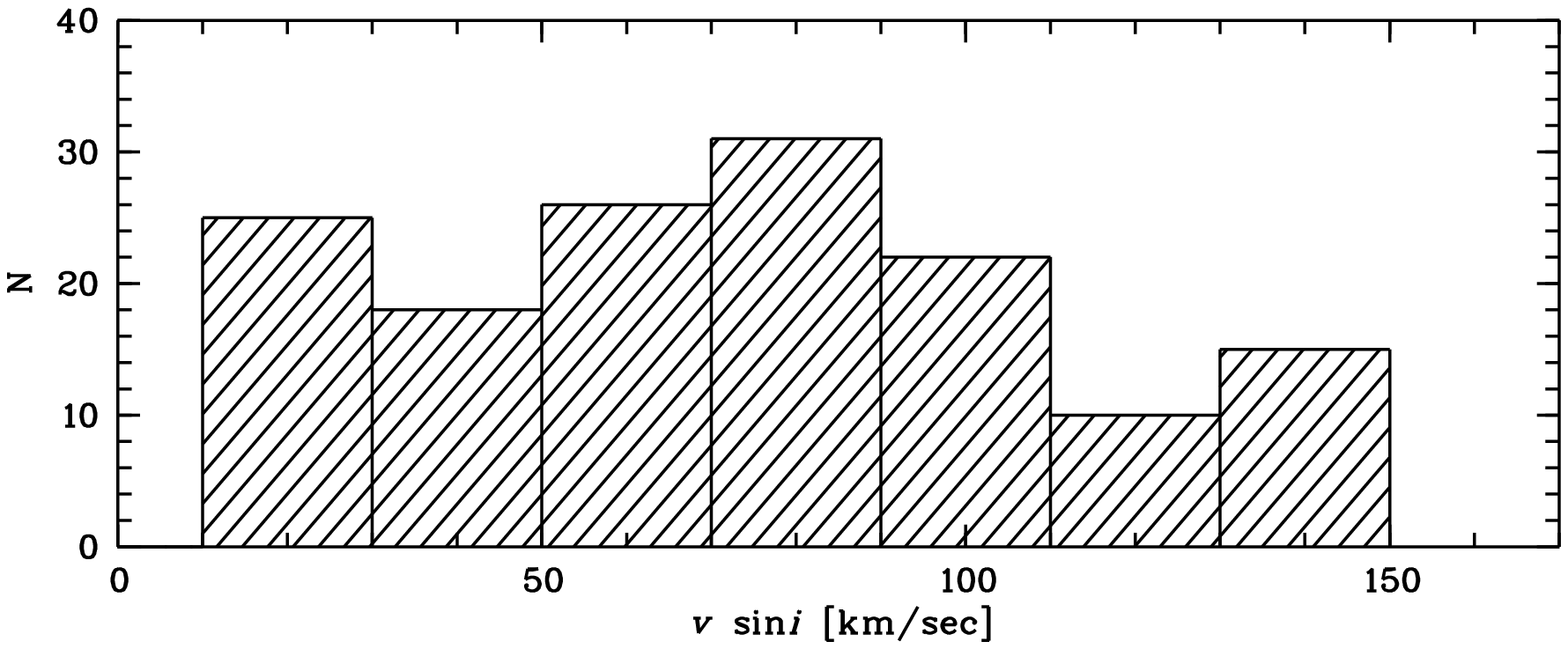}}}
  \mbox{
    \resizebox{.45\hsize}{!}{\includegraphics[bbllx=0,bblly=0,bburx=480,bbury=275]{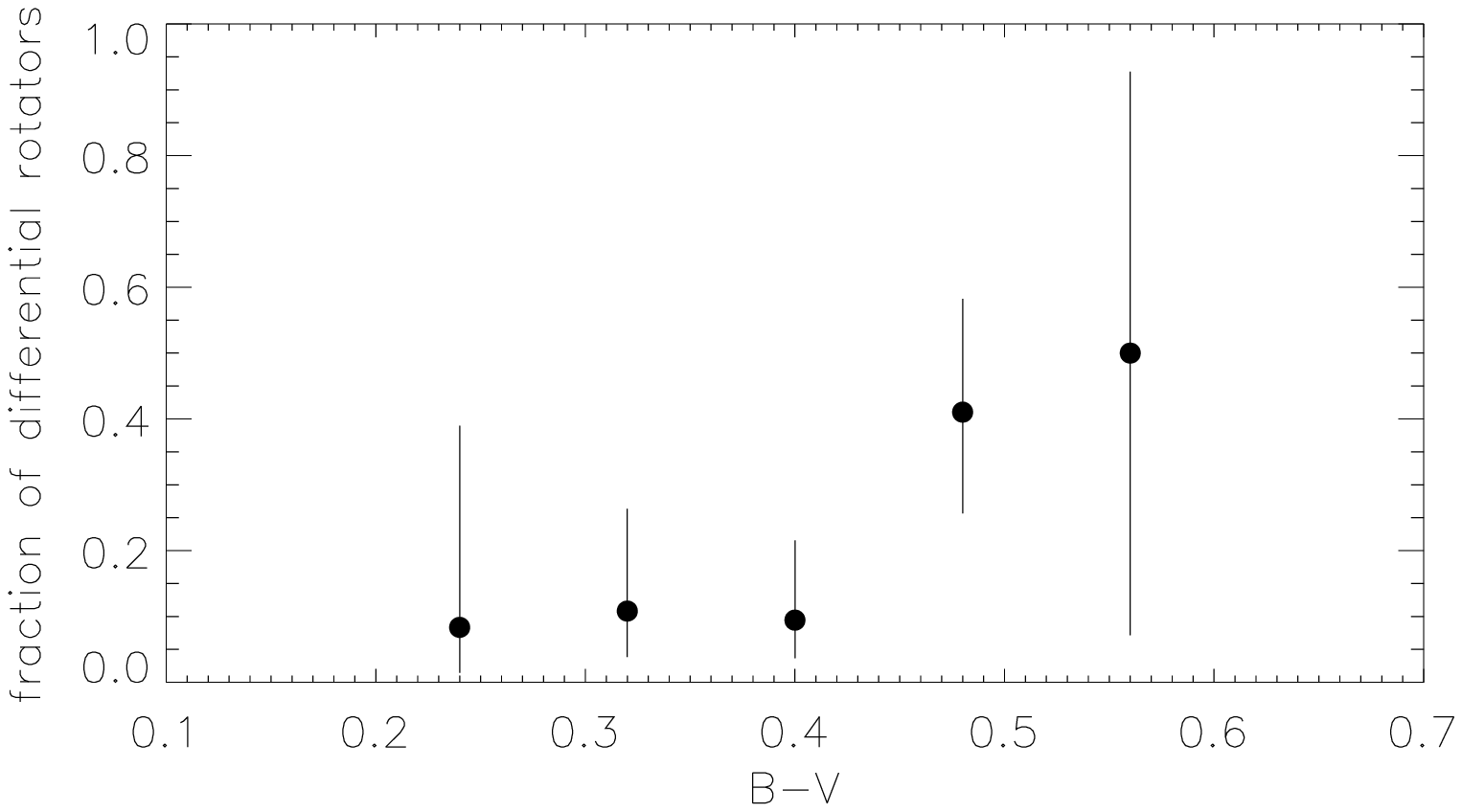}}\hspace{.05\hsize}
    \resizebox{.45\hsize}{!}{\includegraphics[bbllx=0,bblly=0,bburx=480,bbury=275]{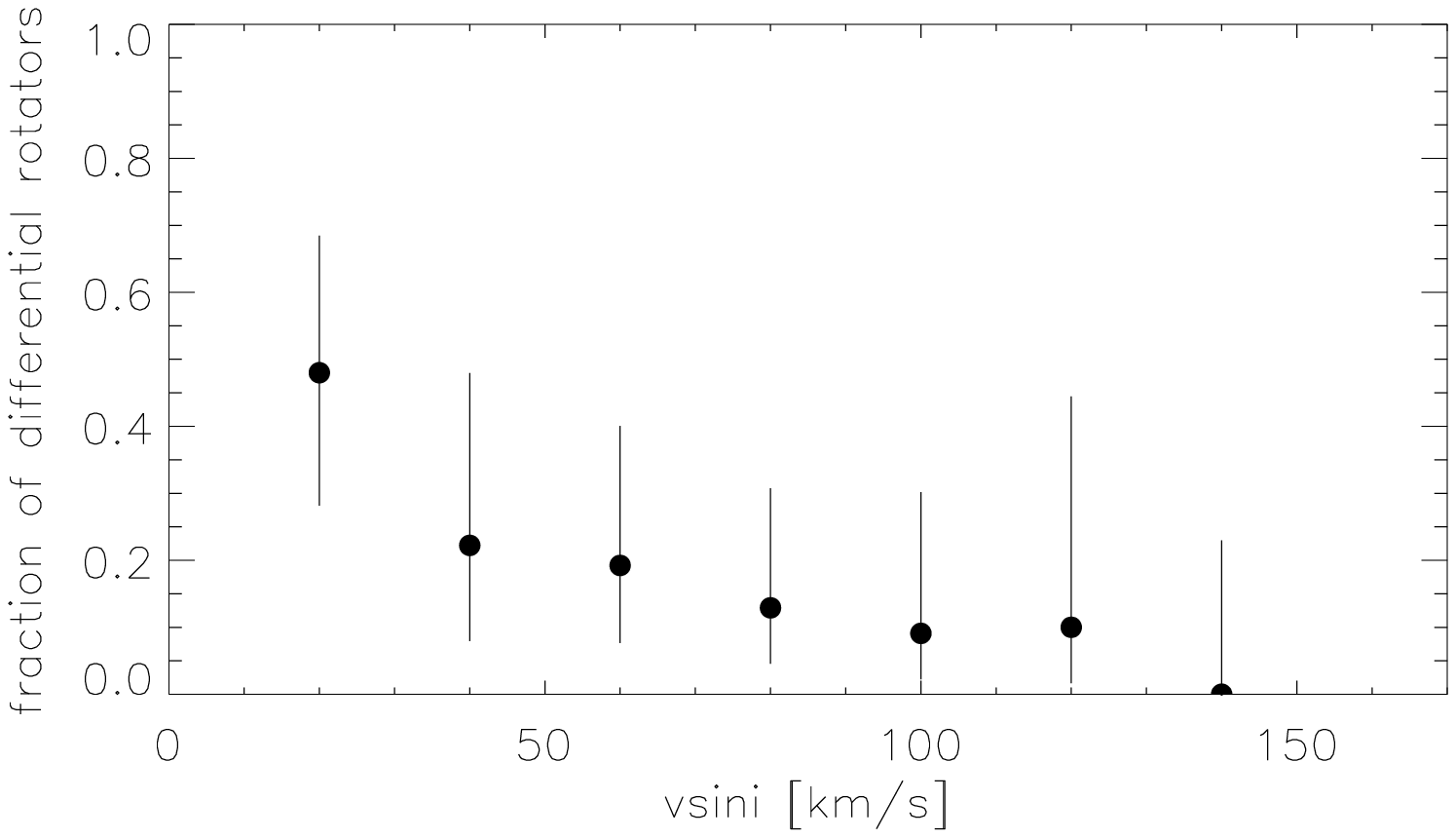}}}
  \caption{\label{plot:histo}Upper panel: Histograms of the sample in $B-V$ (left column)
    and $v\,\sin{i}$ (right column). Lower panel: Fraction of stars
    with strong relative differential rotation in the bins plotted in
    the upper panel.  $2\sigma$-errors are overplotted.}
\end{figure*}

In this section, the hypothesis that the fraction of differentially
rotating stars with $\alpha > 0$ is independent of rotation velocity
$v\,\sin{i}$ and color $B-V$ will be tested. The actual strength of
differential rotation is not taken into account, but will be
investigated in the following sections.

In Fig.\,\ref{plot:vsini_BV}, the whole sample of 147 stars has been
divided into the four arbitrary regions mentioned in
Sect.\,\ref{sect:rotation-color} (dividing at $v\,\sin{i} =
70$\,km\,s$^{-1}$ and $B-V = 0.4$). The number of stars and
differential rotators ($\alpha > 0$, for this sample this means stars
with $\alpha$ above the observational threshold of $\alpha_{\rm min}
\approx 0.05$), and the percentage of differential rotators is given
in Table\,\ref{tab:vs_BV}. The percentage of differentially rotating
stars among slow rotators in regions II and IV is higher than it is
among rapid rotators in regions I and III, respectively; a trend
towards a higher fraction of differential rotators at slower rotation
velocity is visible in both color regimes.  The same is true for the
percentage of differentially rotating stars among late-type stars in
regions III and IV. A comparison to earlier type stars among regions I
and II, yields a higher fraction of differentially rotating stars
towards later spectral type. I tested the hypothesis that subsamples
I--IV are drawn from the same distribution with a total mean of 19\,\%
differential rotators.  Samples II and III are consistent with this
hypothesis (5 and 4 expected differential rotators, respectively). For
samples I and IV, the hypothesis can be rejected at a 99\,\% level
(99.3\,\%, 11 expected, and 99.6\,\%, 8 expected for samples I and IV,
respectively). Thus, the fraction of differential rotators with
$\alpha$ larger than the observational threshold of $\alpha_{\rm min}
\approx 0.05$ -- which does not depend on $v\,\sin{i}$ or color -- is
not constant. It is larger in slower rotators and stars of a later
spectral type.

The distribution of differential rotators in $v\,\sin{i}$ and $B-V$ is
investigated further in Fig.\,\ref{plot:histo}. The upper panel shows
the total number of stars divided into five bins in $B-V$ (left), and
seven bins in $v\,\sin{i}$ (right). The lower panel of
Fig.\,\ref{plot:histo} displays the fraction of differential rotators
($\alpha \ga 0.05$) in the respective color/rotation bins with
$2\sigma$-errors. For example, 48\% of the 25 stars with projected
rotation velocities $v\,\sin{i}$ between 10 and 30\,km\,s$^{-1}$ show
signatures of differential rotation.  Although some bins are sparsely
populated and have large errors, the change from color $B-V = 0.4$ to
$B-V = 0.5$ and the transition from slow rotators to stars with
$v\,\sin{i} > 30$\,km\,s$^{-1}$ are significant.  The trends indicated
in Table\,\ref{tab:vs_BV} stand out in the lower panel of
Fig.\,\ref{plot:histo}. In this sample, profiles with $\alpha \ga
0.05$ are more frequent in slow rotators, which implies that they are
more frequent in stars of later spectral type (due to the sample bias,
cf. Sect.\,\ref{sect:rotation-color}).

\subsection{Differential rotation and rotation period}
\label{sect:P}

\begin{figure*}
  \centering
  \mbox{\resizebox{.45\hsize}{!}{\includegraphics[angle=-90]{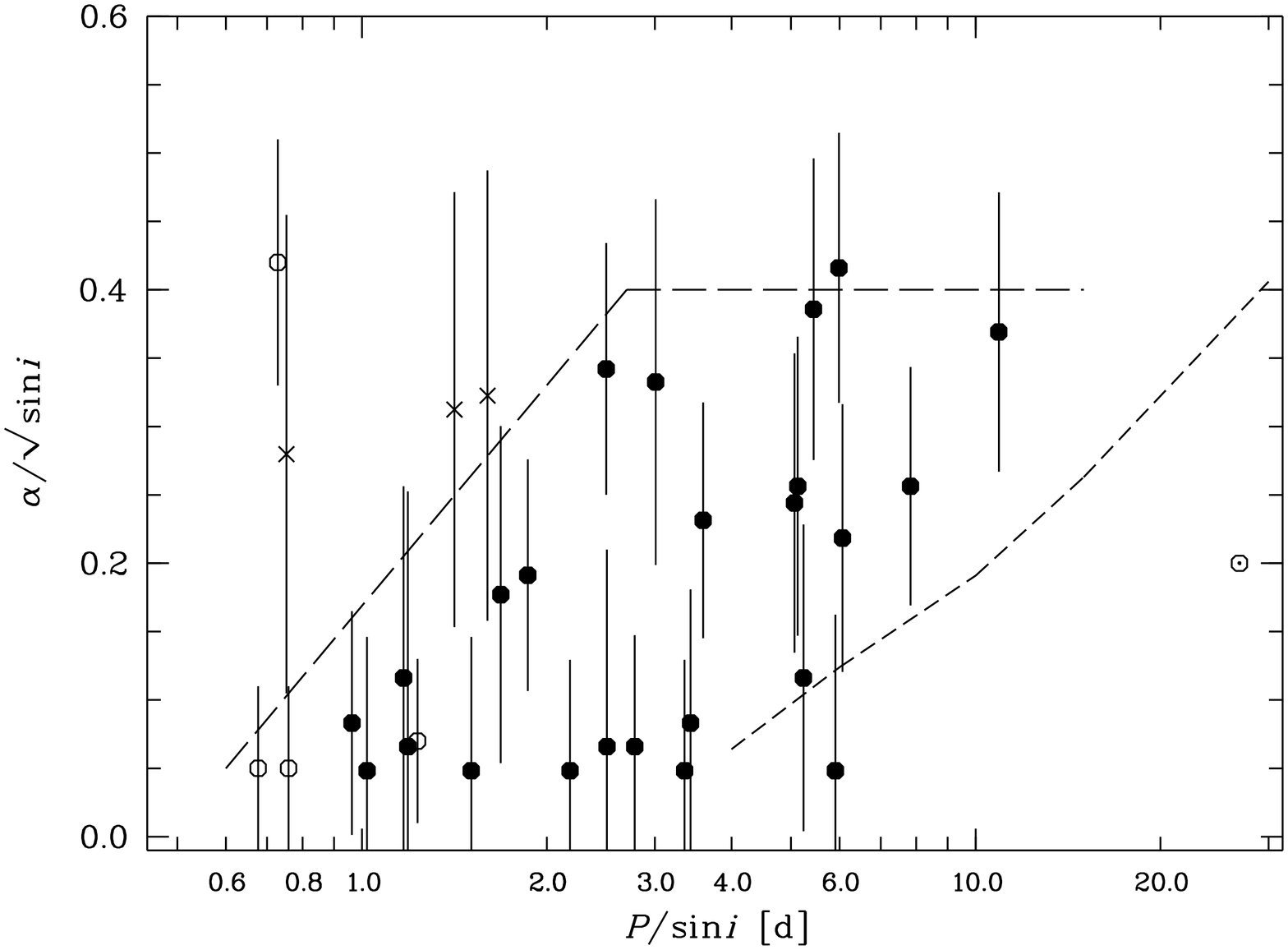}}
    \resizebox{.45\hsize}{!}{\includegraphics[angle=-90]{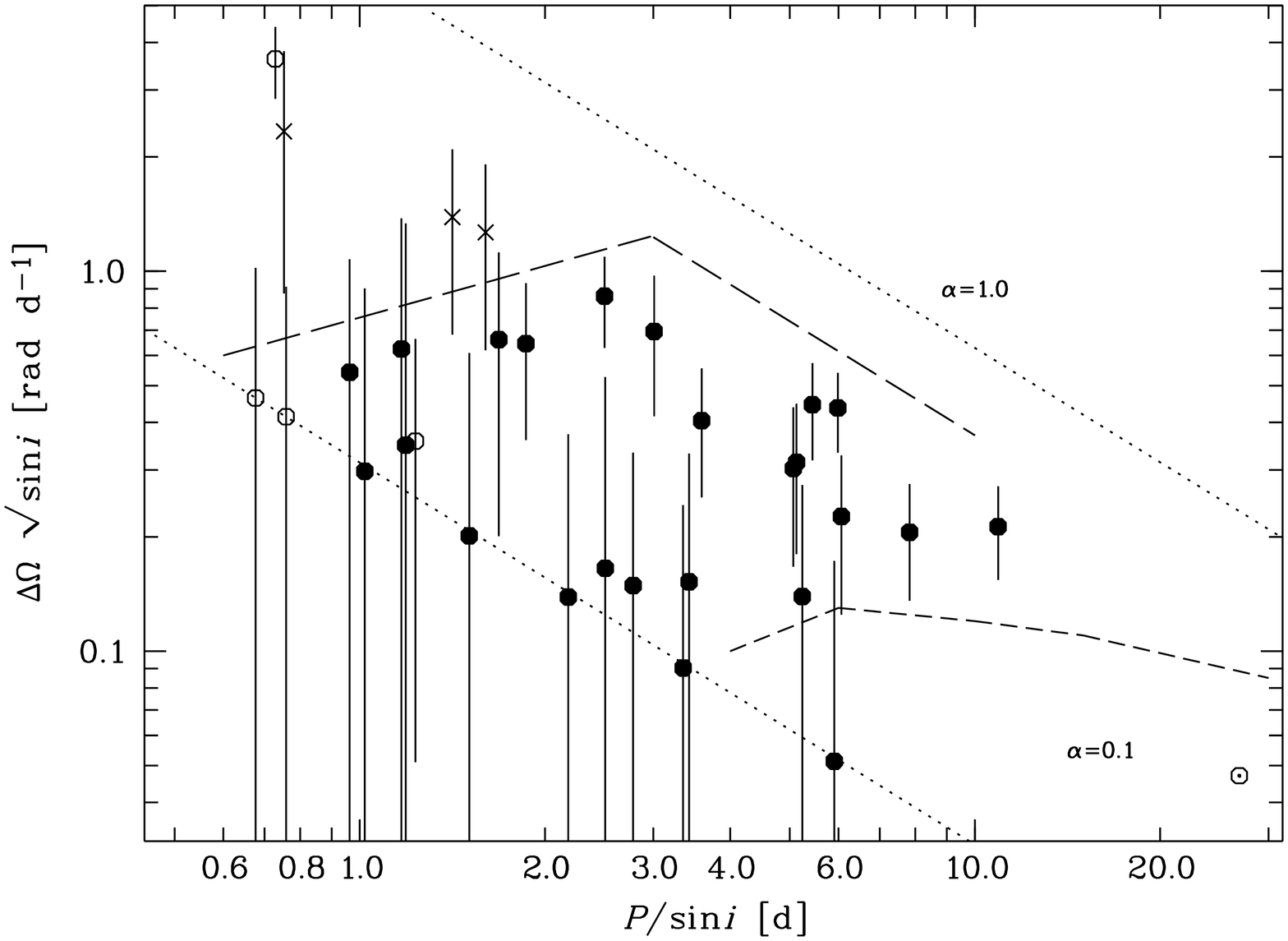}}}
  \caption{\label{plot:dom_P}Differential rotation vs. projected rotation 
    period $P/\sin{i}$. Left: Differential rotation $\alpha =
    \Delta\Omega/\Omega$; right: Absolute shear $\Delta\Omega$ (note
    the logarithmic scale on the ordinate).  Field F-stars are
    indicated by filled circles, open circles display stars from
    cluster observations.  Crosses mark the three differential
    rotators from \cite{Reiners04}. Short dashed lines are models from
    \cite{Kueker05} for an F8-star with a viscosity parameter of
    $c_\nu = 0.15$. The four stars with extraordinarily strong shear
    $\Delta\Omega$ were discussed in Sect.\,\ref{tab:A-stars}, a
    different mechanism is suggested to cause their strong shear. For
    the remaining stars, the upper envelopes are indicated
    qualitatively by the long dashed lines. In the right panel, dotted
    lines show regions of constant differential rotation $\alpha =
    0.1$ (approximate sensitivity of FTM) and $\alpha = 1.0$ (maximum
    if no counter-rotation of the polar regions is allowed for). }
\end{figure*}

In this and the following sections, I will investigate the strength of
differential rotation and especially the maximum strength of
differential rotation at different rotation rates. I discuss
differential rotation in terms of $\alpha = \Delta\Omega/\Omega$ in
Sect.\,\ref{sect:relative}, and analyze the shear $\Delta\Omega$ in
Sect.\,\ref{sect:absolute}. Although both quantities essentially have
the same meaning, it is instructive to look at both of them
separately. Rotation speed will be expressed as a function of
equatorial rotation period instead of rotation velocity. The rotation
period itself is not measured for the majority of stars, so I will use
$P/\sin{i}$ instead, as calculated from measured $v\,\sin{i}$ and the
radius according to Eq.\,\ref{eq:rmbol}.

In the left and right panels of Fig.\,\ref{plot:dom_P}, values of
differential rotation $\alpha$ and absolute shear $\Delta\Omega$ are
plotted against projected rotation period $P/\sin{i}$, respectively.
From the 147 stars in Table\,\ref{tab:observations}, only the 28 stars
with signatures of differential rotation are shown, while the other
119 objects populate the $\alpha = \Delta\Omega/\Omega =0$ region at
$0.5\,{\rm d} < P/\sin{i} < 11.0\,{\rm d}$.  Field stars from this
sample with available $uvby\beta$ measurements are plotted as filled
circles, while the four cluster targets for which the ZAMS-age has
been assumed are plotted as open circles. In both panels, the three
differentially rotating stars from \cite{Reiners04}, which have been
discussed in Sect.\,\ref{sect:Astars}, are marked as crosses.

In the left and right panels of Fig.\,\ref{plot:dom_P}, the long
dashed lines qualitatively indicate the upper envelope of
$\alpha/\sqrt{\sin{i}}$ and $\Delta\Omega\sqrt{\sin{i}}$,
respectively. No fit was attempted, so the lines should only guide the
eye to clarify what will be discussed in the next sections.

\subsubsection{Differential rotation $\alpha$}
\label{sect:relative}

The advantage of using $\alpha$ is that it is measured directly and
its detection does not depend on rotation period, hence radius, while
measuring $\Delta\Omega$ does. Differential rotation $\alpha$ is
smaller than 0.45 for all rotation periods. While a minimum threshold
of $(\alpha/\sqrt{\sin{i}})_{\rm min} \approx 0.05$ applies, the
observational technique has no limitations towards high values of
$\alpha$. Thus, the highest detected value of differential rotation
$(\alpha/\sqrt{\sin{i}})_{\rm max} \approx 0.45$ is not due to
limitations of the FTM.

At rotation periods between two and ten days, the targets populate the
whole region $0 < \alpha/\sqrt{\sin{i}} < 0.45$, while the slower
targets could not be analyzed due to the limitations of the FTM (cp.
Sect\,\ref{sect:method}). Among the rapid rotators with projected
rotation periods that are shorter than two days, the upper envelope
shows a clear decline among the F-stars (filled circles in the left
panel of Fig.\,\ref{plot:dom_P}).  Except for the group of A-stars
discussed in Sect.\,\ref{sect:Astars} (listed in
Table\,\ref{tab:A-stars}), no star with a projected rotation period
less than two days shows $\alpha > 0.2$, and no star with $P/\sin{i} <
1$\,d shows $\alpha > 0.1$.  Neglecting those four stars, the maximum
value in differential rotation, $(\alpha/\sqrt{\sin{i}})_{\rm max}$,
grows from virtually zero at $P = 0.5$\,d to
$(\alpha/\sqrt{\sin{i}})_{\rm max} \approx 0.45$ in stars slower than
$P = 2$\,d. In slower rotators, $(\alpha/\sqrt{\sin{i}})_{\rm max}$
remains approximately constant.

\subsubsection{Absolute shear $\Delta\Omega$}
\label{sect:absolute}

Absolute shear ($\Delta\Omega\sqrt{\sin{i}}$) is shown in the right
panel of Fig.\,\ref{plot:dom_P}. Since observed values of both
$\alpha$ and $\Omega$ depend on inclination $i$, the observed absolute
shear is
\begin{equation} 
  \Delta\Omega_{\rm obs} = \alpha_{\rm obs}\Omega_{\rm obs} = \alpha/\sqrt{\sin{i}}\,\Omega\sin{i} = \Delta\Omega\sqrt{\sin{i}}.
\end{equation}
In this and the following sections, I omit the factor $\sqrt{\sin{i}}$
for readability. Note that in the case of small inclination angles,
the value of $\Delta\Omega$ can be larger than
$\Delta\Omega\sqrt{\sin{i}}$, while $\alpha$ can be smaller than
$\alpha/\sqrt{\sin{i}}$. In the right panel of Fig.\,\ref{plot:dom_P},
the two dotted lines indicate the slopes of constant differential
rotation $\alpha = \Delta\Omega/\Omega$. The upper line is for $\alpha
= 1.0$, i.e., the maximum differential rotation possible regardless of
the observation technique used, and the lower line shows $\alpha =
0.05$, the approximate minimum threshold for the FTM, as explained
above.

As expected from Sect.\,\ref{sect:relative}, the F-stars form a
relatively smooth upper envelope in the maximum absolute shear
$\Delta\Omega_{\rm max}$, as observed at different rotation rates. The
slowest rotators exhibit low values of $\Delta\Omega_{\rm max} \approx
0.2$\,rad\,d$^{-1}$ at $P \approx 10$\,d.  $\Delta\Omega_{\rm max}$
grows towards a faster rotation rate with a maximum between two and
three days before it diminishes slightly with more rapid rotation. The
strongest differential rotation occurs at rotation periods $P$ between
two and three days at a magnitude of $\Delta\Omega \approx
1$\,rad\,d$^{-1}$ (i.e., lapping times on the order of 10\,d).  At
higher rotation rates in the range $0.5\,{\rm d} < P < 2\,{\rm d}$,
the maximum shear has a slope of roughly $\Delta\Omega \propto
P^{+0.4}$. This slope, however, is not constrained well due to the
large uncertainties and sparse sampling. The data are also consistent
with a plateau at $\Delta\Omega \approx 0.7$\,rad\,d$^{-1}$ for $0.5 <
P < 3$\,d. At slower rotation, right from the maximum, the slope is
approximately $\Delta\Omega \propto P^{-1}$.

\cite{Kueker05} recently calculated $\Delta\Omega$ in an F8 star for
different rotation periods. Their results for $\Delta\Omega(P)$ (with
a viscosity coefficient $c_\nu$ = 0.15) are displayed qualitatively in
both panels of Fig.\,\ref{plot:dom_P} as a short-dashed line
\cite[Fig.\,6 in][]{Kueker05}. One of their results is that in their
model $\Delta\Omega$ does not follow a single scaling relation for all
periods \citep[as was approximately the case in the calculations
by][]{Kitchatinov99}, but that a maximum shear arises at a rotation
period of about $P = 7$\,d in the case of the modeled F8-star.
Comparison of their calculations (right panel of
Fig.\,\ref{plot:dom_P}) to the upper envelope suggested in this work
(long-dashed line) still shows a large quantitative discrepancy. The
qualitative slopes of both curves, however, are in reasonable
agreement with each other. The theoretical curve was calculated for an
F8-star.  \cite{Kueker05} also show $\Delta\Omega(P)$ for a solar-type
star, where $\Delta\Omega(P)$ is essentially moved towards higher
rotation periods and lesser shear; i.e. the short-dashed curve in the
right panel of Fig.\,\ref{plot:dom_P} moves to the lower right for
later spectral types. Although earlier spectral types are not
calculated by \cite{Kueker05}, it can be expected that
$\Delta\Omega(P)$ will shift towards higher shear and shorter rotation
periods in stars of an even earlier spectral type. Since most stars
investigated in this sample are earlier than F8, this would suggest
qualitative consistency between theoretical curves and the slope of
$\Delta\Omega_{\rm max}$ shown here.

\subsubsection{Comparing different techniques}
\label{sect:Barnes}

Differential rotation measurements are now available from a variety of
observational techniques (see Sect.\,\ref{sect:introduction}),
comparison of results from techniques becomes possible. However, such
a comparison has to be carried out with great care. Photometrically
measured periods, for example, are only sensitive to latitudes covered
by spots; they reflect only parts of the rotation law and are always
lower limits. Furthermore, temperature has been shown to be the
dominating factor for the strength of differential rotation
\citep{Kitchatinov99, Kueker05, Barnes05}, which has to be taken into
account when analyzing the rotation dependence of differential
rotation.

In the past, analyses of relations between rotation and differential
rotation have generally assumed a monotonic scaling relation between
period $P$ (or angular velocity $\Omega = 2\pi/P$) and $\Delta
\Omega$. Such a relation was expected from calculations by
\cite{Kitchatinov99}. As mentioned above, \cite{Kueker05} recently
presented new calculations showing that the $\Delta
\Omega$\,vs.\,$\Omega$-relation may have a temperature dependent
maximum.

Searching for dependence on angular velocity, \cite{Barnes05} recently
compiled data from differential rotation measurements from DI,
photometric monitoring, and FTM. Fitting a single power law to the
compiled data, they derive $\Delta\Omega \propto
\Omega^{+0.15\pm0.10}$, which is compared to the case of a G2 dwarf
calculated in \cite{Kitchatinov99}. From the latter, they cite the
theoretical G2 dwarf relation as $\Delta\Omega \propto
\Omega^{+0.15}$, and claim agreement to their fit.  Although the work
of \cite{Kitchatinov99} has been superseded by \cite{Kueker05}, it
should be mentioned that \cite{Kitchatinov99} report $\Delta\Omega
\propto \Omega^{-0.15}$, implying stronger shear for slower rotation
instead of weaker \citep[note that ][also report a negative exponent
for their solar-like star model at periods less than 20\,d]{Kueker05}.
In fact, the large scatter in the compilation of all measurements from
different techniques \citep[Fig.\,3 in ][]{Barnes05} and the severe
bias due to systematic uncertainties (like the lack of small values
$\Delta\Omega$ in the sample measured with FTM) leads to any
conclusion about the period dependence from such a heterogeneous
sample very uncertain. The new data in this paper does not
significantly improve this situation and no analysis that improves
upon the one performed by \cite{Barnes05} can be expected.

Although the constantly growing amount of differential rotation
measurements provides a relatively large sample, the results from
analyzing all measurements as one sample do not yet provide convincing
evidence for a unique rotation dependence of differential rotation
over the whole range of rotation periods.

\subsection{Upper bound of differential rotation and effective temperature} 

\begin{figure}
  \centering
  \resizebox{\hsize}{!}{\includegraphics[angle=-90]{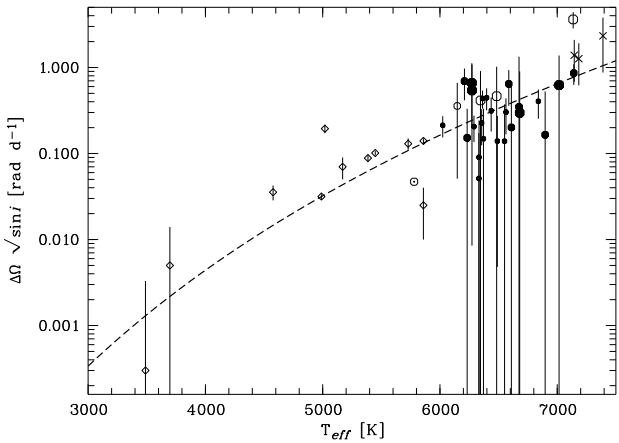}}
  \caption{\label{plot:dom_Teff}Differential rotation $\Delta\Omega$
    with effective temperature $T_{\rm eff}$ for sample stars (symbols
    as in Fig.\,\ref{plot:dom_P}). Big circles indicate large rotation
    velocity. Stars analyzed with DI are indicated as squares, and the
    fit to DI targets from \cite{Barnes05} is overplotted. F-stars
    with the strongest differential rotation are consistent with the
    fit. }
\end{figure}

Effective temperature is the second parameter after rotation to govern
stellar differential rotation, since convection zone depth, as well as
convection velocity, are very sensitive to $T_{\rm eff}$.
\cite{Kueker05} and \cite{Kitchatinov99} report stronger differential
rotation with higher effective temperatures when comparing a
solar-type star to stars of spectral types F8 and K5. The differential
rotators in the sample of F-stars investigated here span a range in
effective temperature between 6000\,K and 7150\,K. As a result,
analysis of temperature effects on differential rotation is limited by
the small range of targets in $T_{\rm eff}$, and is biased by the
large range in rotation periods, as discussed above. Thus, I limit the
analysis of temperature dependence to a comparison to differential
rotation in stars that are significantly cooler than $T_{\rm eff} =
6000$\,K.

Considering stars with effective temperatures in the range $3400\,{\rm
  K} < T_{\rm eff} < 6000$\,K, \cite{Barnes05} found a power-law
dependence in their measurements of absolute shear $\Delta\Omega$ on
stellar surface temperature. Their result is compared to the sample of
this work in Fig.\,\ref{plot:dom_Teff}. Absolute shear follows the
trend expected by \cite{Kueker05} and \cite{Kitchatinov99} with a
stronger surface shear at higher effective temperatures. Although a
large scatter is observed in the F-star measurements, they
qualitatively follow this trend and connect to cooler stars at roughly
the expected values. In addition, the rotation rate of the F-stars is
indicated by symbol size, larger symbols displaying higher rotation
rate. Stars exhibiting a shear in excess of the expected rate for
their temperature tend to show very rapid rotation.  Thus, the
investigated F-stars agree qualitatively with the temperature fit
derived by \cite{Barnes05}. A more quantitative analysis is
complicated by the large scatter among measurements of absolute shear,
which is visible also in the sample of \cite{Barnes05}.  Again, it
should be noted that the values plotted for the F-star sample only
display stars for which signatures of differential rotation have been
measured. A large number of stars populate the region of weaker
surface shear or rigid rotation ($\alpha = \Delta\Omega = 0$), and the
temperature law applies only to the strongest differential rotators.

\section{Conclusions}

The sample of stars with rotation laws measured from spectral
broadening profiles with the FTM is constantly growing. In this work,
44 new observations of stars of spectral type F and later were added
to the results from former publications. Currently, rotation laws have
been analyzed in a homogeneous data set of 147 stars of spectral type
F and later, and in a second data set in 78 stars of spectral type A.
Among all these observations covering the temperature range between
5600\,K and 10\,000\,K \citep[including A-stars from][]{Reiners04}, 31
stars exhibit signatures of solar-like differential rotation. Only
three of them are of spectral type A. In the HR-diagram, differential
rotators appear near and on the cool side of the convection boundary.
No differentially rotating star hotter than 7400\,K is known, and it
is obvious that the signatures of solar-like differential rotation are
closely connected to the existence of deep convective envelopes.

Most differential rotators can be found near to the ZAMS at young
ages, but due to the limited sample and severe selection effects, this
needs confirmation from a less biased sample.  Particularly, a number
of slower rotators with temperatures around $T_{\rm eff} = 6600$\,K
are needed.

Four differential rotators are very close to the convection boundary.
All four show extraordinarily strong absolute shear and exhibit
projected rotation velocities within 10\% around $v\,\sin{i} =
100$\,km\,s$^{-1}$.  It is suggested that these stars form a group of
objects in which rotation velocity and convection zone depth
facilitate very strong absolute shear and that the mechanism causing
the shear is different from the later F-type stars.

Among the F-stars, differential rotation occurs in the whole range of
temperatures and rotation rates. The sample of 147 stars of spectral
type F and later was investigated for the dependence of differential
rotation on rotation and temperature. 28 of them (19\%) exhibit
signatures of differential rotation. The distribution of differential
rotation was approached with two different strategies: (i)
investigation of the fraction of stars exhibiting differential
rotation ($\alpha > 0$); and (ii) analysis of the maximum $\alpha$ and
maximum $\Delta\Omega$ as a function of rotation period and
temperature. The first approach reflects the typical values of
$<$$\alpha$$>$ and $<$$\Delta\Omega$$>$ at any given temperature and
rotation rate, while the second focuses on the question how strong the
absolute shear can possibly be in such stars.  Due to the large
uncertainties in the measurements and the high minimum threshold in
differential rotation, the mean values of $\alpha$ and $\Delta\Omega$
are not particularly meaningful.  Furthermore, it is not clear whether
a smooth transition from stars exhibiting strong differential rotation
to ``rigidly'' rotating stars ($\alpha < 0.05$) or a distinction
between these two groups exists, and the mentioned approaches were
preferred for the analysis.

In the sample, hotter stars generally rotate more rapidly, and effects
due to rotation velocity and temperature cannot be disentangled.  The
distribution of differential rotators depends on color and/or rotation
rate, and the fraction of stars with differential rotation ($\alpha >
0$) increases with cooler temperature and/or slower rotation. It is
not clear what this means to the mean shear, $<$$\Delta\Omega$$>$. For
example, it is not inconsistent with $<$$\Delta\Omega$$>$ being
constant at all rotation rates.  In this case, $<$$\alpha$$> =
<$$\Delta\Omega/\Omega$$>$ would be smaller in more rapidly rotating
stars, and thus a lower fraction of stars would exhibit differential
rotation $\alpha$ above the observational threshold, as is observed.

On the other hand, the maximum observed values of differential
rotation, $\alpha_{\rm max}$, and of the absolute shear,
$\Delta\Omega_{\rm max}$, do vary depending on the rotation rate. The
strongest absolute shear of $\Delta\Omega \approx 1$\,rad\,d$^{-1}$ is
found at rotation periods between two and three days with
significantly smaller values in slower rotators. The more rapidly
rotating stars show a slight decrease in absolute shear as well,
although the sparse data are also consistent with a plateau at
$\Delta\Omega \approx 0.7$\,rad\,d$^{-1}$ for $0.5 < P < 3$\,d. A
maximum in differential rotation $\Delta\Omega$ has recently been
predicted by \cite{Kueker05} for an F8-star, although of lesser
strength and at a slower rotation rate, a difference that may in parts
be due to their later spectral type.

The investigated sample does not cover a wide range in effective
temperature since only very few late-type field stars rotate fast
enough for the method applied. Although temperature is expected to
strongly influence the strength of differential rotation, the large
range in rotation rate and the connection between rotation rate and
temperature in the sample makes conclusions about temperature effects
insecure.  The results were compared to differential rotation
measurements in cooler stars and found in qualitative agreement with
an extrapolation of the empirical temperature dependence that
\cite{Barnes05} found when analyzing a sample of differentially rotation
measurements done with DI.

The implications of the relations discussed here for stellar magnetic
activity and the nature of the dynamo working in F-type stars still
remain unclear from an observational point of view. Naively, one would
expect stronger magnetic activity to occur in stars with stronger
differential rotation among groups of comparable temperature or
rotation rate. Those stars for which X-ray measurements are available,
however, do not yet exhibit such a trend, but a meaningful
investigation is hampered by the limited amount of data points
available (especially for comparable temperature or rotation
velocity). It has been shown that qualitative conclusions can be
derived from the currently available measurements of stellar rotation
laws, but a more detailed investigation of the consequences on the
dynamo operating in F-type stars has to wait until a statistically
better-defined sample of stars is available.

\begin{acknowledgements}
  I am thankful to G. Basri for carefully reading the manuscript and
  for very helpful discussions, and to J. Schmitt for valuable
  comments on an earlier version of the manuscript. I thank the
  referee, Dr. John Barnes, for a careful and very constructive
  report.  A.R. has received research funding from the European
  Commission's Sixth Framework Programme as an Outgoing International
  Fellow (MOIF-CT-2004-002544).
\end{acknowledgements}

\appendix

\section{The first direct comparison to Doppler Imaging}
\label{sect:DI}

\begin{figure}
  \centering
  \resizebox{\hsize}{!}{\includegraphics[angle=-90]{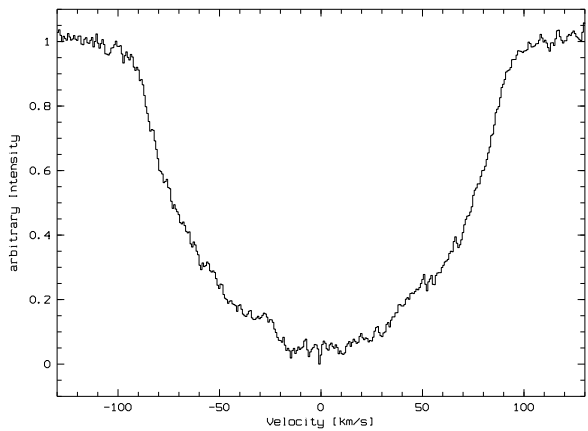}}
  \caption{\label{plot:HD307938}Broadening function of HD~307\,938. This
    object has also been studied with Doppler Imaging by
    \cite{Marsden05}. }
\end{figure}

Recently, \cite{Marsden05} have presented Doppler Images of
HD~307\,938, a young active G dwarf in IC\,2602. This star was also
observed for this project in the FLAMES/UVES campaign, and I report on
the rotation law in Table\,\ref{tab:uvesstars}.  \cite{Marsden05} took
a time series over four nights detecting spectroscopic variability.
Their data is contaminated by a significant amount of sunlight
reflected by the moon, which they have carefully removed before
constructing Doppler Images. The result of Doppler Imaging is that
HD~307\,938 has a cool polar cap extending down to $\sim 60\degr$
latitude, and $v\,\sin{i} = 92 \pm 0.5$\,km\,$^{-1}$, which
excellently corresponds to $v\,\sin{i} = 93.7 \pm 4.7$\,km\,s$^{-1}$,
as derived here (on comments about uncertainties in $v\,\sin{i}$, see
Sect.\,\ref{sect:results}).  They also report marginal differential
rotation of $\Delta\Omega = 0.025 \pm 0.015$\,rad\,d$^{-1}$ with a
$1\sigma$ error. From this result, one expects a value of $q_2/q_1$
that is only marginally less than 1.76, the value for rigid rotation
and the best guess for the limb darkening parameter. The cool polar
spot found on HD~307\,938 influences $q_2/q_1$ as well, as it enlarges
$q_2/q_1$ much more than the small deviation from rigid rotation does;
$q_2/q_1$ is thus expected to be larger than 1.76. The spectrum
secured for the analysis presented here is not contaminated by
sunlight, and the broadening function derived is shown in
Fig.\,\ref{plot:HD307938}.  The profile does not reveal large
asymmetry, although a spot may be visible around $v \approx
-30$\,km\,s$^{-1}$. The profile is fully consistent with the
broadening functions presented in \cite{Marsden05}, who were able to
show temporal variations in the profile from lower quality data. The
profile parameter determined from FTM is $q_2/q_1 = 1.78 \pm 0.01$,
indicating no signs of solar-like differential rotation that is large
enough to be detected with this method. However, the fact that
$q_2/q_1$ is slightly larger than 1.76 supports the idea that a cool
spot occupies the polar caps. Thus, the finding of marginal
differential rotation and a cool polar spot is consistent with the
result derived from FTM (besides the good consistency in
$v\,\sin{i}$). This is the first time that a direct comparison of the
results is possible, since the majority of Doppler Imaging targets
usually show spot signatures that are stronger than what can be dealt
with using FTM.

\section{Tables of stars with measured differential rotation.}
\label{sect:appendix}

\begin{table*}
  \caption[]{Field Stars with measured differential rotation.}
  \label{tab:fieldstars}
  \begin{tabular}{crrrrrccccc}
    \hline
    \hline
    \noalign{\smallskip}
    Star & HR & $v\,\sin{i}$  & $¸\delta~v\,\sin{i}$ & $q_{\rm 2}/q_{\rm 1}$ & $\delta q_{\rm 2}/q_{\rm 1}$ & $\Delta\Omega$ & $\delta\Delta\Omega$ & $T_{\rm eff}$ & M$_V$ & $P/\sin{i}$\\
    & & [km\,s$^{-1}$] & [km\,s$^{-1}$] & & & rad d$^{-1}$ & rad d$^{-1}$ & K & mag & d\\
    \noalign{\smallskip}
    \hline
    \noalign{\smallskip}
    \object{HD   432}  &    21  &   71.0  &   3.6 & 1.78 & 0.03 & 0.00 &       &  6763 & 0.96 & 3.08 \\ 
\object{HD  4089}  &   187  &   23.5  &   1.2 & 1.82 & 0.02 & 0.00 &       &  6161 & 2.75 & 5.01 \\ 
\object{HD  4247}  &   197  &   42.7  &   2.1 & 1.77 & 0.04 & 0.00 &       &  6825 & 3.08 & 1.90 \\ 
\object{HD  4757}  &   230  &   93.8  &   4.7 & 1.73 & 0.03 & 0.00 &       &  6706 & 1.34 & 2.00 \\ 
\object{HD  6706}  &   329  &   46.0  &   2.3 & 1.83 & 0.06 & 0.00 &       &  6551 & 2.77 & 2.21 \\ 
\object{HD  6903}  &   339  &   87.7  &   4.4 & 1.69 & 0.02 & 0.54 & 0.534 &  6273 & 3.38 & 0.96 \\ 
\object{HD 15524}  &   728  &   59.8  &   3.0 & 1.81 & 0.08 & 0.00 &       &  6592 & 2.44 & 1.95 \\ 
\object{HD 17094}  &   813  &   45.1  &   2.3 & 1.49 & 0.02 & 0.86 & 0.231 &  7141 & 2.17 & 2.50 \\ 
\object{HD 17206}  &   818  &   25.6  &   1.3 & 1.70 & 0.02 & 0.15 & 0.184 &  6371 & 3.68 & 2.78 \\ 
\object{HD 18256}  &   869  &   17.2  &   0.9 & 1.71 & 0.04 & 0.05 & 0.121 &  6332 & 2.93 & 5.91 \\ 
\object{HD 22001}  &  1083  &   13.2  &   0.7 & 1.83 & 0.07 & 0.00 &       &  6601 & 2.98 & 6.88 \\ 
\object{HD 22701}  &  1107  &   55.0  &   2.8 & 1.83 & 0.03 & 0.00 &       &  6610 & 2.70 & 1.88 \\ 
\object{HD 23754}  &  1173  &   13.8  &   0.7 & 1.87 & 0.05 & 0.00 &       &  6518 & 2.98 & 6.75 \\ 
\object{HD 24357}  &  1201  &   65.8  &   3.3 & 1.82 & 0.05 & 0.00 &       &  6895 & 2.83 & 1.35 \\ 
\object{HD 25457}  &  1249  &   18.0  &   0.9 & 1.71 & 0.02 & 0.09 & 0.152 &  6333 & 4.07 & 3.35 \\ 
\object{HD 25621}  &  1257  &   16.7  &   0.8 & 1.73 & 0.03 & 0.00 &       &  6091 & 2.26 & 9.02 \\ 
\object{HD 27459}  &  1356  &   78.3  &   3.9 & 1.78 & 0.05 & 0.00 &       &  7642 & 2.15 & 1.29 \\ 
\object{HD 28677}  &  1432  &  134.8  &   6.7 & 1.77 & 0.03 & 0.00 &       &  6981 & 2.76 & 0.67 \\ 
\object{HD 28704}  &  1434  &   88.1  &   4.4 & 1.79 & 0.03 & 0.00 &       &  6672 & 2.43 & 1.30 \\ 
\object{HD 29875}  &  1502  &   47.8  &   2.4 & 1.82 & 0.06 & 0.00 &       &  7080 & 3.39 & 1.37 \\ 
\object{HD 29992}  &  1503  &   97.5  &   4.9 & 1.75 & 0.04 & 0.00 &       &  6742 & 2.64 & 1.04 \\ 
\object{HD 30034}  &  1507  &  103.6  &   5.2 & 1.82 & 0.06 & 0.00 &       &  7484 & 2.34 & 0.92 \\ 
\object{HD 30652}  &  1543  &   17.3  &   0.9 & 1.78 & 0.03 & 0.00 &       &  6408 & 3.59 & 4.23 \\ 
\object{HD 33167}  &  1668  &   47.5  &   2.4 & 1.82 & 0.03 & 0.00 &       &  6493 & 2.10 & 2.97 \\ 
\object{HD 35296}  &  1780  &   15.9  &   0.8 & 1.75 & 0.02 & 0.00 &       &  6060 & 4.11 & 4.09 \\ 
\object{HD 37147}  &  1905  &  109.9  &   5.5 & 1.77 & 0.07 & 0.00 &       &  7621 & 2.25 & 0.88 \\ 
\object{HD 41074}  &  2132  &   87.8  &   4.4 & 1.78 & 0.06 & 0.00 &       &  6912 & 2.45 & 1.20 \\ 
\object{HD 43386}  &  2241  &   19.5  &   1.0 & 1.83 & 0.03 & 0.00 &       &  6512 & 3.50 & 3.77 \\ 
\object{HD 44497}  &  2287  &   89.4  &   4.5 & 1.79 & 0.03 & 0.00 &       &  7010 & 1.79 & 1.56 \\ 
\object{HD 46273}  &  2384  &  106.9  &   5.3 & 1.74 & 0.05 & 0.00 &       &  6674 & 2.10 & 1.25 \\ 
\object{HD 48737}  &  2484  &   66.1  &   3.3 & 1.78 & 0.04 & 0.00 &       &  6496 & 2.32 & 1.93 \\ 
\object{HD 51199}  &  2590  &   91.7  &   4.6 & 1.77 & 0.04 & 0.00 &       &  6730 & 2.06 & 1.45 \\ 
\object{HD 55052}  &  2706  &   81.8  &   4.1 & 1.80 & 0.04 & 0.00 &       &  6668 & 0.63 & 3.21 \\ 
\object{HD 56986}  &  2777  &  129.7  &   6.5 & 1.75 & 0.08 & 0.00 &       &  6837 & 1.95 & 1.05 \\ 
\object{HD 57927}  &  2816  &   89.5  &   4.5 & 1.75 & 0.02 & 0.00 &       &  6772 & 1.40 & 1.99 \\ 
\object{HD 58579}  &  2837  &  147.1  &   7.4 & 1.81 & 0.03 & 0.00 &       &  7053 & 1.69 & 0.98 \\ 
\object{HD 58946}  &  2852  &   59.0  &   3.0 & 1.76 & 0.07 & 0.00 &       &  6892 & 2.79 & 1.54 \\ 
\object{HD 60111}  &  2887  &  117.5  &   5.9 & 1.80 & 0.05 & 0.00 &       &  7181 & 3.01 & 0.65 \\ 
\object{HD 61035}  &  2926  &  124.2  &   6.2 & 1.80 & 0.03 & 0.00 &       &  6986 & 3.09 & 0.62 \\ 
\object{HD 61110}  &  2930  &   91.1  &   4.6 & 1.78 & 0.04 & 0.00 &       &  6575 & 1.51 & 1.98 \\ 
\object{HD 62952}  &  3015  &  127.5  &   6.4 & 1.80 & 0.08 & 0.00 &       &  6933 & 1.53 & 1.26 \\ 
\object{HD 64685}  &  3087  &   47.6  &   2.4 & 1.88 & 0.03 & 0.00 &       &  6838 & 3.04 & 1.73 \\ 
\object{HD 67483}  &  3184  &   52.4  &   4.1 & 1.50 & 0.07 & 0.69 & 0.279 &  6209 & 2.07 & 3.01 \\ 
\object{HD 69548}  &  3254  &   53.9  &   2.7 & 1.77 & 0.05 & 0.00 &       &  6705 & 3.53 & 1.27 \\ 
\object{HD 70958}  &  3297  &   45.5  &   2.3 & 1.74 & 0.05 & 0.00 &       &  6230 & 3.53 & 1.76 \\ 
\object{HD 70958}  &  3297  &   46.1  &   2.3 & 1.75 & 0.01 & 0.00 &       &  6230 & 3.53 & 1.74 \\ 
\object{HD 72943}  &  3394  &   56.8  &   2.8 & 1.70 & 0.06 & 0.17 & 0.361 &  6897 & 1.81 & 2.51 \\ 
\object{HD 75486}  &  3505  &  128.1  &   6.4 & 1.79 & 0.02 & 0.00 &       &  6993 & 1.22 & 1.42 \\ 
\object{HD 76143}  &  3537  &   83.0  &   4.2 & 1.79 & 0.03 & 0.00 &       &  6579 & 1.58 & 2.10 \\ 
\object{HD 76582}  &  3565  &   90.5  &   4.5 & 1.80 & 0.05 & 0.00 &       &  7884 & 2.58 & 0.87 \\ 
\object{HD 77370}  &  3598  &   60.4  &   3.0 & 1.71 & 0.03 & 0.20 & 0.408 &  6609 & 2.97 & 1.51 \\ 
\object{HD 77601}  &  3603  &  140.7  &   7.0 & 1.82 & 0.02 & 0.00 &       &  6421 & 0.44 & 2.21 \\ 
\object{HD 79940}  &  3684  &  117.2  &   5.9 & 1.78 & 0.03 & 0.00 &       &  6397 & 0.88 & 2.18 \\ 
\object{HD 81997}  &  3759  &   30.4  &   1.5 & 1.73 & 0.01 & 0.00 &       &  6471 & 3.28 & 2.72 \\ 
\object{HD 82554}  &  3795  &  129.7  &   6.5 & 1.63 & 0.05 & 0.66 & 0.460 &  6272 & 1.32 & 1.68 \\ 
\object{HD 83287}  &  3829  &  102.5  &   5.1 & 1.77 & 0.08 & 0.00 &       &  7815 & 2.69 & 0.74 \\ 
\object{HD 83962}  &  3859  &  140.3  &   7.0 & 1.72 & 0.08 & 0.00 &       &  6507 & 1.53 & 1.30 \\ 
\object{HD 84607}  &  3879  &   93.1  &   4.7 & 1.79 & 0.01 & 0.00 &       &  7000 & 1.78 & 1.51 \\ 
\object{HD 88215}  &  3991  &   97.5  &   4.9 & 1.78 & 0.05 & 0.00 &       &  6823 & 3.04 & 0.85 \\ 
\object{HD 89254}  &  4042  &   63.5  &   3.2 & 1.81 & 0.02 & 0.00 &       &  7173 & 2.42 & 1.57 \\ 

    \noalign{\smallskip}
    \hline
  \end{tabular}
\end{table*}

\addtocounter{table}{-1}
\begin{table*}
  \caption[]{continued}
  \begin{tabular}{crrrrrccccc}
    \hline
    \hline
    \noalign{\smallskip}
    Star & HR & $v\,\sin{i}$  & $¸\delta~v\,\sin{i}$ & $q_{\rm 2}/q_{\rm 1}$ & $\delta q_{\rm 2}/q_{\rm 1}$ & $\Delta\Omega$ & $\delta\Delta\Omega$ & $T_{\rm eff}$ & M$_V$  & $P/\sin{i}$\\
    & & [km\,s$^{-1}$] & [km\,s$^{-1}$] & & & rad d$^{-1}$ & rad d$^{-1}$ & K & mag & d\\
    \noalign{\smallskip}
    \hline
    \noalign{\smallskip}
    \object{HD 89449}  &  4054  &   17.3  &   1.7  & 1.44 & 0.04 & 0.45 & 0.127 &  6398 & 3.05 & 5.44 \\ 
\object{HD 89569}  &  4061  &   12.2  &   0.7  & 1.57 & 0.02 & 0.21 & 0.070 &  6290 & 3.10 & 7.83 \\ 
\object{HD 89571}  &  4062  &  133.9  &   6.7  & 1.79 & 0.07 & 0.00 &       &       &      &      \\ 
\object{HD 90089}  &  4084  &   56.2  &   2.8  & 1.70 & 0.09 & 0.35 & 0.987 &  6674 & 3.60 & 1.19 \\ 
\object{HD 90589}  &  4102  &   51.6  &   2.6  & 1.93 & 0.04 & 0.00 &       &  6794 & 2.95 & 1.68 \\ 
\object{HD 96202}  &  4314  &   93.4  &   4.7  & 1.77 & 0.05 & 0.00 &       &  6747 & 2.34 & 1.25 \\ 
\object{HD 99329}  &  4410  &  137.9  &   6.9  & 1.73 & 0.04 & 0.00 &       &  6990 & 2.66 & 0.68 \\ 
\object{HD100563}  &  4455  &   13.5  &   0.7  & 1.67 & 0.04 & 0.14 & 0.134 &  6489 & 3.61 & 5.24 \\ 
\object{HD105452}  &  4623  &   23.5  &   1.2  & 1.59 & 0.02 & 0.40 & 0.151 &  6839 & 2.98 & 3.60 \\ 
\object{HD106022}  &  4642  &   77.2  &   3.9  & 1.86 & 0.06 & 0.00 &       &  6651 & 2.28 & 1.60 \\ 
\object{HD107326}  &  4694  &  132.2  &   6.6  & 1.80 & 0.05 & 0.00 &       &  7107 & 2.08 & 0.90 \\ 
\object{HD108722}  &  4753  &   97.0  &   4.8  & 1.78 & 0.04 & 0.00 &       &  6490 & 1.71 & 1.74 \\ 
\object{HD109085}  &  4775  &   60.0  &   3.0  & 1.75 & 0.05 & 0.00 &       &  6813 & 3.13 & 1.32 \\ 
\object{HD109141}  &  4776  &  135.7  &   6.8  & 1.79 & 0.03 & 0.00 &       &  6881 & 2.84 & 0.66 \\ 
\object{HD110385}  &  4827  &  105.2  &   5.3  & 1.75 & 0.01 & 0.00 &       &  6717 & 1.71 & 1.49 \\ 
\object{HD110834}  &  4843  &  133.3  &   6.7  & 1.73 & 0.06 & 0.00 &       &  6244 & 0.99 & 1.93 \\ 
\object{HD111812}  &  4883  &   63.0  &   3.2  & 1.78 & 0.05 & 0.00 &       &  5623 & 2.87 & 2.22 \\ 
\object{HD112429}  &  4916  &  119.6  &   6.0  & 1.77 & 0.03 & 0.00 &       &  7126 & 2.97 & 0.66 \\ 
\object{HD114378}  &  4968  &   19.9  &   1.0  & 1.76 & 0.01 & 0.00 &       &  6324 & 3.82 & 3.40 \\ 
\object{HD115810}  &  5025  &   99.2  &   5.0  & 1.82 & 0.06 & 0.00 &       &  7185 & 1.95 & 1.24 \\ 
\object{HD116568}  &  5050  &   36.8  &   1.8  & 1.73 & 0.01 & 0.00 &       &  6485 & 3.20 & 2.32 \\ 
\object{HD118889}  &  5138  &  140.6  &   7.0  & 1.82 & 0.06 & 0.00 &       &  6951 & 2.40 & 0.76 \\ 
\object{HD119756}  &  5168  &   63.9  &   3.2  & 1.78 & 0.03 & 0.00 &       &  6809 & 3.06 & 1.29 \\ 
\object{HD120136}  &  5185  &   15.6  &   1.0  & 1.57 & 0.04 & 0.31 & 0.134 &  6437 & 3.38 & 5.13 \\ 
\object{HD121370}  &  5235  &   13.5  &   1.3  & 1.46 & 0.03 & 0.21 & 0.059 &  6024 & 2.36 & 0.90 \\ 
\object{HD122066}  &  5257  &   40.6  &   2.0  & 1.81 & 0.01 & 0.00 &       &  6395 & 2.18 & 3.45 \\ 
\object{HD124780}  &  5337  &   70.7  &   3.5  & 1.80 & 0.04 & 0.00 &       &  7204 & 2.21 & 1.54 \\ 
\object{HD124850}  &  5338  &   15.0  &   0.8  & 1.91 & 0.04 & 0.00 &       &  6075 & 2.85 & 7.71 \\ 
\object{HD127739}  &  5434  &   55.8  &   2.8  & 1.81 & 0.01 & 0.00 &       &  6787 & 2.20 & 2.20 \\ 
\object{HD127821}  &  5436  &   55.6  &   2.8  & 1.75 & 0.04 & 0.00 &       &  6601 & 3.76 & 1.14 \\ 
\object{HD129153}  &  5473  &  105.7  &   5.3  & 1.78 & 0.06 & 0.00 &       &  7693 & 2.70 & 0.73 \\ 
\object{HD129502}  &  5487  &   47.0  &   2.4  & 1.80 & 0.03 & 0.00 &       &  6695 & 2.94 & 1.91 \\ 
\object{HD129926}  &  5497  &  112.5  &   5.6  & 1.74 & 0.03 & 0.00 &       &  6048 & 3.91 & 0.64 \\ 
\object{HD132052}  &  5570  &  113.2  &   5.7  & 1.77 & 0.06 & 0.00 &       &  6964 & 2.19 & 1.03 \\ 
\object{HD136359}  &  5700  &   20.3  &   1.0  & 1.78 & 0.01 & 0.00 &       &  6296 & 3.02 & 4.86 \\ 
\object{HD136751}  &  5716  &   72.7  &   3.6  & 1.77 & 0.02 & 0.00 &       &  6810 & 2.31 & 1.60 \\ 
\object{HD138917}  &  5788  &   85.8  &   5.4  & 1.99 & 0.06 & 0.00 &       &       &      &      \\ 
\object{HD139225}  &  5804  &  104.4  &   5.2  & 1.76 & 0.03 & 0.00 &       &  6960 & 2.50 & 0.97 \\ 
\object{HD139664}  &  5825  &   71.6  &   3.6  & 1.77 & 0.05 & 0.00 &       &  6681 & 3.57 & 0.94 \\ 
\object{HD142908}  &  5936  &   75.8  &   3.8  & 1.80 & 0.05 & 0.00 &       &  6849 & 2.33 & 1.50 \\ 
\object{HD143466}  &  5960  &  141.3  &   7.1  & 1.77 & 0.05 & 0.00 &       &  7235 & 2.20 & 0.77 \\ 
\object{HD147365}  &  6091  &   72.5  &   3.6  & 1.77 & 0.06 & 0.00 &       &  6657 & 3.46 & 0.99 \\ 
\object{HD147449}  &  6093  &   76.4  &   3.8  & 1.79 & 0.04 & 0.00 &       &  6973 & 2.70 & 1.21 \\ 
\object{HD147449}  &  6093  &   77.1  &   3.9  & 1.80 & 0.03 & 0.00 &       &  6973 & 2.70 & 1.20 \\ 
\object{HD148048}  &  6116  &   84.8  &   4.2  & 1.78 & 0.05 & 0.00 &       &  6731 & 2.43 & 1.32 \\ 
\object{HD150557}  &  6205  &   61.8  &   3.1  & 1.83 & 0.04 & 0.00 &       &  6959 & 2.01 & 2.06 \\ 
\object{HD151613}  &  6237  &   47.5  &   2.4  & 1.84 & 0.07 & 0.00 &       &  6630 & 2.80 & 2.06 \\ 
\object{HD155103}  &  6377  &   57.9  &   2.9  & 1.81 & 0.05 & 0.00 &       &  7150 & 2.55 & 1.63 \\ 
\object{HD156295}  &  6421  &  107.4  &   5.4  & 1.79 & 0.07 & 0.00 &       &  7818 & 2.56 & 0.75 \\ 
\object{HD160915}  &  6595  &   12.4  &   0.6  & 1.60 & 0.03 & 0.23 & 0.101 &  6356 & 3.58 & 6.07 \\ 
\object{HD164259}  &  6710  &   69.3  &   3.5  & 1.75 & 0.04 & 0.00 &       &  6704 & 2.70 & 1.44 \\ 
\object{HD165373}  &  6754  &   79.9  &   4.0  & 1.74 & 0.03 & 0.00 &       &  6976 & 2.08 & 1.54 \\ 
\object{HD171834}  &  6987  &   71.3  &   3.6  & 1.77 & 0.04 & 0.00 &       &  6622 & 2.50 & 1.58 \\ 
\object{HD173417}  &  7044  &   53.9  &   2.7  & 1.83 & 0.04 & 0.00 &       &  6780 & 1.81 & 2.74 \\ 
\object{HD173667}  &  7061  &   18.0  &   2.0  & 1.40 & 0.02 & 0.44 & 0.104 &  6363 & 2.78 & 5.99 \\ 
\object{HD175317}  &  7126  &   17.1  &   0.9  & 1.58 & 0.04 & 0.30 & 0.136 &  6563 & 3.11 & 5.07 \\ 
\object{HD175824}  &  7154  &   53.7  &   2.7  & 1.69 & 0.03 & 0.15 & 0.179 &  6232 & 1.72 & 3.43 \\ 
\object{HD182640}  &  7377  &   87.3  &   4.4  & 1.67 & 0.06 & 0.62 & 0.753 &  7016 & 2.46 & 1.17 \\ 
\object{HD185124}  &  7460  &   87.0  &   4.4  & 1.71 & 0.03 & 0.30 & 0.604 &  6680 & 2.98 & 1.02 \\ 
\object{HD186005}  &  7489  &  149.9  &   7.5  & 1.79 & 0.04 & 0.00 &       &  6988 & 1.68 & 0.98 \\ 

    \noalign{\smallskip}
    \hline
  \end{tabular}
\end{table*}

\addtocounter{table}{-1}
\begin{table*}
  \caption[]{continued}
  \begin{tabular}{crrrrrccccc}
    \hline
    \hline
    \noalign{\smallskip}
    Star & HR & $v\,\sin{i}$  & $¸\delta~v\,\sin{i}$ & $q_{\rm 2}/q_{\rm 1}$ & $\delta q_{\rm 2}/q_{\rm 1}$ & $\Delta\Omega$ & $\delta\Delta\Omega$ & $T_{\rm eff}$ & M$_V$  & $P/\sin{i}$\\
    & & [km\,s$^{-1}$] & [km\,s$^{-1}$] & & & rad d$^{-1}$ & rad d$^{-1}$ & K & mag & d\\
    \noalign{\smallskip}
    \hline
    \noalign{\smallskip}
    \object{HD187532}  &  7553 &   77.5  &   3.9  & 1.75 & 0.03  & 0.00 &       &  6788 & 3.34  & 0.94 \\ 
\object{HD189245}  &  7631 &   72.6  &   3.6  & 1.74 & 0.03  & 0.00 &       &  6259 & 4.06  & 0.85 \\ 
\object{HD190004}  &  7657 &  136.1  &   6.8  & 1.80 & 0.04  & 0.00 &       &  6974 & 2.47  & 0.76 \\ 
\object{HD197692}  &  7936 &   41.7  &   2.1  & 1.62 & 0.02  & 0.65 & 0.286 &  6587 & 3.33  & 1.86 \\ 
\object{HD199260}  &  8013 &   13.7  &   0.7  & 1.79 & 0.03  & 0.00 &       &  6213 & 4.18  & 4.36 \\ 
\object{HD201636}  &  8099 &   58.8  &   2.9  & 1.86 & 0.03  & 0.00 &       &  6700 & 2.20  & 2.15 \\ 
\object{HD203925}  &  8198 &   70.7  &   3.5  & 1.80 & 0.03  & 0.00 &       &  6845 & 1.15  & 2.77 \\ 
\object{HD205289}  &  8245 &   57.5  &   2.9  & 1.74 & 0.03  & 0.00 &       &  6525 & 3.24  & 1.44 \\ 
\object{HD206043}  &  8276 &  134.0  &   6.7  & 1.74 & 0.04  & 0.00 &       &  7092 & 2.87  & 0.62 \\ 
\object{HD207958}  &  8351 &   69.3  &   3.5  & 1.84 & 0.03  & 0.00 &       &  6747 & 2.97  & 1.26 \\ 
\object{HD210302}  &  8447 &   13.6  &   0.7  & 1.72 & 0.04  & 0.00 &       &  6465 & 3.52  & 5.46 \\ 
\object{HD210459}  &  8454 &  147.4  &   7.4  & 1.76 & 0.03  & 0.00 &       &  6376 & 0.21  & 2.89 \\ 
\object{HD213051}  &  8558 &   48.4  &   2.4  & 1.85 & 0.08  & 0.00 &       &       &       &      \\ 
\object{HD213845}  &  8592 &   35.7  &   1.8  & 1.71 & 0.02  & 0.14 & 0.234 &  6551 & 3.35  & 2.18 \\ 
\object{HD219693}  &  8859 &   19.9  &   1.0  & 1.81 & 0.01  & 0.00 &       &  6461 & 2.82  & 5.14 \\ 
\object{HD220657}  &  8905 &   73.4  &   3.7  & 1.74 & 0.03  & 0.00 &       &  5801 & 2.41  & 2.15 \\ 

    \noalign{\smallskip}
    \hline
  \end{tabular}
\end{table*}

\begin{table*}
  \caption[]{\label{tab:uvesstars}Stars in cluster fields, FLAMES/UVES observations}
  \begin{tabular}{ccrrrrccccc}
    \hline
    \hline
    \noalign{\smallskip}
    Star & Cluster & $v\,\sin{i}$  & $¸\delta~v\,\sin{i}$ & $q_{\rm 2}/q_{\rm 1}$ & $\delta q_{\rm 2}/q_{\rm 1}$ & $\Delta\Omega$ & $\delta\Delta\Omega$ & $T_{\rm eff}$ & M$_V$  & $P/\sin{i}$\\
    && [km\,s$^{-1}$] & [km\,s$^{-1}$] & & & rad d$^{-1}$ & rad d$^{-1}$ & K & mag & d\\
    \noalign{\smallskip}
    \hline
    \noalign{\smallskip}
    \object{NGC 6475 69}         & NGC\,6475 &  94.9  &   4.7  & 1.76 & 0.01  & 0.00 &       &  6822 & 3.27 &  0.75 \\ 
\object{NGC 6475 41}         & NGC\,6475 &  80.1  &   4.0  & 1.71 & 0.01  & 0.41 & 0.497 &  6346 & 3.86 &  0.76 \\ 
\object{BD+20 2161}          & Praesepe  &  72.7  &   3.6  & 1.76 & 0.01  & 0.00 &       &  6864 & 3.23 &  0.99 \\ 
\object{BD+20 2170}          & Praesepe  &  94.4  &   4.7  & 1.71 & 0.01  & 0.46 & 0.557 &  6484 & 3.69 &  0.68 \\ 
\object{Cl* NGC 2632 KW 230} & Praesepe  &  78.9  &   3.9  & 1.75 & 0.01  & 0.00 &       &  6628 & 3.51 &  0.85 \\ 
\object{Cl* IC 2391 L 33}    & IC\,2391  &  81.1  &   4.1  & 1.74 & 0.01  & 0.00 &       &  6414 & 3.78 &  0.77 \\ 
\object{HD 307938}           & IC\,2602  &  93.7  &   4.7  & 1.78 & 0.01  & 0.00 &       &  5743 & 4.64 &  0.51 \\ 
\object{HD 308012}           & IC\,2602  &  45.7  &   2.3  & 1.70 & 0.01  & 0.36 & 0.306 &  6147 & 4.12 &  1.23 \\ 
\object{Cl* IC 4665 V 69}    & IC\,4665  &  45.1  &   2.3  & 1.74 & 0.01  & 0.00 &       &  6180 & 4.08 &  1.26 \\ 
\object{Cl* IC 4665 V 102}   & IC\,4665  & 105.0  &  12.0  & 1.39 & 0.01  & 3.62 & 0.776 &  7136 & 2.93 &  0.73 \\ 
\object{Cl* IC 4665 V 97}    & IC\,4665  &  52.0  &   2.6  & 1.90 & 0.01  & 0.00 &       &  6628 & 3.51 &  1.29 \\ 

    \noalign{\smallskip}
    \hline
  \end{tabular}
\end{table*}



\begin{thebibliography}{} 
  
\bibitem[Balthasar et al., 1986]{Balthasar86} Balthasar, H.,
  V\'azques, M., \& W\"ohl, H., 1986, \aap, 185, 87
  
\bibitem[Barnes et al., 2005]{Barnes05}Barnes, J.R., Collier Cameron,
  A., Donati, J.-F., James, D.J., Marsden, S.C.,~\& Petit, P., 2005,
  \mnras, 357, L1

\bibitem[Bruning, 1981]{Bruning81}Bruning, D.H., 1981, \apj, 248, 271
  
\bibitem[Domiciano de Souza et al., 2004]{Domiciano04}Domiciano de
  Souza, A., Zorec, J., Jankov, S., Vakili, F., Abe, L., \&
  Janot-Pacheco, E., 2004, \aap, 418, 781

\bibitem[Donahue et al., 1996]{Donahue96}Donahue, R.A., Saar, S.H., \&
  Baliunas, S.L., 1996, \apj, 466, 384
  
\bibitem[Donati \& Collier Cameron, 1997]{Donati97} Donati J.-F.,
  Collier Cameron A., 1997, MNRAS 291, 1
  
\bibitem[Dravins et al., 1990]{Dravins90} Dravins D., Lindegren L.,
  Torkelsson U., 1990, A\&A, 237, 137

\bibitem[Gizon \& Solanki, 2004]{Gizon04}Gizon, L., \& Solanki, S.K.,
  2004, \solphys, 220, 169

\bibitem[Gray, 1976]{Gray76} Gray D.F., 1976, The observation and
  analysis of stellar photospheres, Wiley, New York

\bibitem[Gray, 1977]{Gray77}Gray, D.F., 1977, \apj, 211, 198
  
\bibitem[Gray, 1984]{Gray84}Gray, D.F., 1984, \apj, 277, 640

\bibitem[Gray, 1998]{Gray98}Gray, D.F., 1998, ASP Conf. Ser. 154, 10th
  Cambridge Workshop on Cool Stars, Stellar Systems and the Sun
  
\bibitem[Gray \& Nagel, 1989]{Gray89}Gray, D.F., \& Nagel, T., 1989,
  \apj, 341, 421
  
\bibitem[Hall, 1991]{Hall91}Hall, D.S., 1991, in The Sun and the cool
  Stars, Springer Verlag, New York
  
\bibitem[Hayes, 1978]{Hayes78}Hayes, D.S., 1978, IAU Symp. 80, eds.
  Philip, A.G.D., \& Hayes, D.S., p. 65

\bibitem[Hauck \& Mermilliod, 1998]{Hauck98} Hauck B., Mermilliod M.,
  1998, A\&AS, 129, 431
  
\bibitem[Kitchatinov \& R\"udiger, 1999]{Kitchatinov99}Kitchatinov,
  L.L, and R\"udiger, G., 1999, A\&A, 344, 911
  
\bibitem[K\"uker \& R\"udiger, 2005]{Kueker05}K\"uker, M. \&
  R\"udiger, G., 2005, \aap, 422, 1023
  
\bibitem[Marcy, 1984]{Marcy84}Marcy, G.W., 1984, \apj, 276, 286
  
\bibitem[Marsden et al., 2005]{Marsden05}Marsden, S.C., Waite, I.A.,
  Carter, B.D., \& Donati, J.-F., 2005, MNRAS, 359, 711

\bibitem[Moon, 1985]{Moon85a} Moon T.T., 1985, Coom.
  Univ. London Obs. 78

\bibitem[Moon \& Dworetsky, 1985]{Moon85b} Moon T.T., Dworetsky M.M.,
  1985, MNRAS, 217, 305

\bibitem[Napiwotzki et al., 1993]{Napiwotzki93} Napiwotzki R.,
  Sch\"onberner D., and Wenske V., 1993, A\&A, 268, 653
  
\bibitem[Petit et al., 2002]{Petit02}Petit, P., Donati, J.-F., \&
  Collier Cameron, A., 2002, MNRAS, 334, 374
  
\bibitem[Reiners, 2003]{Reiners03} Reiners, A., 2003, \aap, 408, 707
  
\bibitem[Reiners \& Schmitt, 2002a]{Reiners02a} Reiners, A., \&
  Schmitt, J.H.M.M., 2002a, \aap, 384, 155
  
\bibitem[Reiners \& Schmitt, 2002b]{Reiners02b}Reiners, A., \&
  Schmitt, J.H.M.M., 2002b, \aap, 388, 1120

\bibitem[Reiners \& Schmitt, 2003a]{Reiners03a} Reiners, A.,\& Schmitt,
  J.H.M.M., 2003a, \aap, 398, 647
  
\bibitem[Reiners \& Schmitt, 2003b]{Reiners03b} Reiners, A., \& Schmitt,
  J.H.M.M., 2003b, \aap, 412, 813
  
\bibitem[Reiners \& Royer, 2004]{Reiners04} Reiners, A., \& Royer, F.,
  2004, \aap, 415, 325
  
\bibitem[Reiners et al., 2005]{Reiners05} Reiners, A., H\"unsch, M.,
  Hempel, M., \& Schmitt, J.H.M.M., 2005, \aap, 430, L17
  
\bibitem[Schrijver \& Title, 2001]{Schrijver01}Schrijver, C.J., \&
  Title, A.M., 2001, \apj, 551, 1099
  
\bibitem[Schrijver \& Zwaan, 2000]{Schrijver00}Schrijver, C.J., \&
  Zwaan, C,, 2000, Solar and Magnetic Activity, Cambridge University
  Press
  
\bibitem[Schou, 1998]{Schou98}Schou, J., 1998, \apj, 122, 293

\bibitem[Siess et al., 2000]{Siess00} Siess L., Dufour E., and
  Forestini M., 2000, A\&A, 358, 593
  
\bibitem[Solanki, 1991]{Solanki91}Solanki, S.K., 1991, RvMA, 4, 208
  
\bibitem[Stahler \& Palla, 2004]{Stahler04}Stahler, S.W., \& Palla,
  F., 2004, The Formation of Stars, Wiley-VCH

\bibitem[Johns-Krull \& Valenti, 1996]{Johns-Krull96}Johns-Krull,
  C.M., \& Valenti, J.A., 1996, \apj, 459, 95

\bibitem[Wolter et al., 2005]{Wolter05}Wolter, U., Schmitt, J.H.M.M.,
  \& and Van Wyk, F., 2005, \aap, 436, 261

\end{thebibliography}
\end{document}